\documentclass[australian,english,prl, singlespace, twocolumn]{revtex4-1}
\usepackage[T1]{fontenc}
\usepackage[latin9]{inputenc}
\usepackage{color}
\usepackage{array}
\usepackage{float}
\usepackage{multirow}
\usepackage{amsmath}
\usepackage{amssymb}
\usepackage{graphicx}

\makeatletter

\newcommand*\LyXThinSpace{\,\hspace{0pt}}
\providecommand{\tabularnewline}{\\}

\@ifundefined{definecolor}
 {\usepackage{color}}{}
\makeatother

\makeatother

\usepackage{babel}
\begin{document}

\title{Leggett-Garg tests of macro-realism for dynamical cat-states evolving
in a nonlinear medium}

\author{Manushan Thenabadu$^{1}$ and M. D. Reid$^{1,3}$}

\affiliation{$^{1}$Centre for Quantum and Optical Science Swinburne University
of Technology, Melbourne, Australia}

\affiliation{$^{3}$Institute of Theoretical Atomic, Molecular and Optical Physics
(ITAMP),Harvard University, Cambridge, Massachusetts, USA}
\begin{abstract}
We show violations of Leggett-Garg inequalities to be possible for
single-mode cat-states evolving dynamically in the presence of a nonlinear
quantum interaction arising from, for instance, a Kerr medium. In
order to prove the results, we derive a generalised version of the
Leggett-Garg inequality involving different cat-states at different
times. The violations demonstrate failure of the premise of macro-realism
as defined by Leggett and Garg, provided extra assumptions associated
with experimental tests are valid. With the additional assumption
of stationarity, violations of the Leggett-Garg inequality are predicted
for the multi-component cat-states observed in the Bose-Einstein condensate
and superconducting circuit experiments of Greiner et al. {[}Nature
\textbf{419}, 51, (2002){]} and Kirchmair et al. {[}Nature, \textbf{495},
205 (2013){]}. The violations demonstrate a mesoscopic quantum coherence,
by negating that the system can be in a classical mixture of mesoscopically
distinct coherent states. Higher orders of nonlinearity are also studied
and shown to give strong violation of Leggett-Garg inequalities.
\end{abstract}
\maketitle

\section{Introduction}

\textcolor{black}{In quantum mechanics, the Schrodinger cat-state
is of interest because it is a superposition of two macroscopically
distinct states \cite{s-cat}. This gives a paradox, because the Copenhagen
interpretation of a system in such a superposition is that the system
cannot be regarded as being in one state or the other, prior to a
measurement. The assumption that a macroscopic system is in one or
other of two macroscopically distinct states prior to measurement
is referred to as ``macroscopic realism''. Cat-states have been
created in laboratories \cite{cat-states-review,collapse-revival-bec,collapse-revival-super-circuit,cat-states,supercond-microwave-cats}.
However, cat-states are usually signified by evidence that the system
is in a superposition, rather than a classical mixture, of the two
states. This evidence is presented within the framework of quantum
mechanics. }

\textcolor{black}{In 1985, Leggett and Garg were motivated to test
macroscopic realism directly, without assumptions based on the validity
of quantum mechanics \cite{legggarg}. This gives the potential for
a stronger demonstration of a cat-state. Leggett and Garg originally
considered a dynamical system that is always found, by some measurement,
to be in one of two macroscopically distinct states at any given time
e.g. Schrodinger's cat is always found to be dead or alive. They derived
inequalities which if violated negated the validity of a form of macroscopic
realism, commonly referred to as ``}\textcolor{black}{\emph{macro-realism}}\textcolor{black}{''.
Macro-realism involves an additional premise, called macroscopic noninvasive
measurability. The beauty of the Leggett-Garg inequality is that (similar
to Bell inequalities \cite{bell-1}) a whole class of classical hidden
variable theories can be potentially falsified by an experiment. }

\textcolor{black}{There has been experimental evidence for violations
of Leggett-Garg inequalities \cite{experiment-lg,lgexpphotonweak,emeryreview,NSTmunro,small-systemlg,stat-1zhou,stat-2-neutrino,atom-lg}.
Many of the tests and proposals to date however have not addressed
macroscopic states, some exceptions being Refs. \cite{NSTmunro,atom-lg,jordan_kickedqndlg2}
which address systems such as a superconducting qubit, or a single
atom. There have also been proposals given for tests of macro-realism
using macroscopic or mesoscopic atomic systems \cite{laura-lg-two-well,massiveosci,Mitchell,bogdan-two-well}.
One of the most commonly considered cat-states is the superposition
of two single-mode coherent states, given as
\begin{equation}
|\psi\rangle=c_{1}|\alpha_{1}\rangle+c_{2}|\alpha_{2}\rangle\label{eq:cat2-1}
\end{equation}
($c_{i}$ are complex amplitudes) where $|\alpha\rangle$ is a coherent
state \cite{milburn-holmes,yurke-stoler}. This cat-state has been
successfully created in a single mode microwave field, with a 100
photon separation ($|\alpha_{1}-\alpha_{2}|^{2}\sim100$) between
the two distinct states, using a superconducting circuit \cite{supercond-microwave-cats,collapse-revival-super-circuit}.
This is one of the largest cat-states ever created in a laboratory,
with a record quantifiable macroscopic quantum coherence \cite{cat-states-review}.
Similar cat-states have been created in a Bose-Einstein condensate
\cite{collapse-revival-bec}, giving the potential to test macro-realism
for many atoms. To the best of our knowledge however, a Leggett-Garg
test for these cat-state systems has not yet been proposed.}

\textcolor{black}{In this paper, we give such a test. We show how
to test for Leggett-Garg's macro-realism for cat-states of the type
(\ref{eq:cat2-1}). In fact, we will consider the multi-component
cat-states $|\psi\rangle=\sum_{i}c_{i}|\alpha_{i}\rangle$, defined
as a superposition of }\textcolor{black}{\emph{multiple }}\textcolor{black}{coherent
states. Here, we assume the phase-space separation of the coherent
states $|\alpha_{i}\rangle$ and $|\alpha_{j}\rangle$ is large i.e.
$|\alpha_{i}-\alpha_{j}|\rightarrow\infty$. In this way, we give
an experimental proposal that relates directly to the cat-state experiments
of Greiner et al. \cite{collapse-revival-bec} and Kirchmair et al.
\cite{collapse-revival-super-circuit}, thus giving the possibility
of strong tests of macro-realism involving large micro-wave cat-states,
and cat-states with a large number of atoms.}

\textcolor{black}{We study the specific case considered by Yurke and
Stoler \cite{yurke-stoler}, where a single oscillator or field mode
prepared in a coherent state undergoes a nonlinear interaction described
by the Hamiltonian
\begin{equation}
H_{NL}=\Omega\hat{n}^{k}\label{eq:hnl}
\end{equation}
($k>1$). It is well known that the interaction (\ref{eq:hnl}) leads
to the formation of cat-states \cite{milburn-holmes,collapse-revival-bec,collapse-revival-super-circuit,yurke-stoler,wrigth-walls-gar}.
The details of the dynamical evolution depend on whether $k$ is odd
or even. For simplicity in this paper, we focus only on even $k$,
and consider the cases $k=2$ and $k>2$ separately. Cat-states are
also formed for odd $k$ \cite{yurke-stoler} however, and the techniques
presented in this paper may well be useful for this case. Here $\Omega$
is the strength of the nonlinearity and $\hat{n}=\hat{a}^{\dagger}\hat{a}$
is the number operator, where $\hat{a}$, $\hat{a}^{\dagger}$ are
the bosonic destruction and annihilation operators. We show in this
paper that, by considering three successive times of evolution, one
can violate a Leggett-Garg inequality based on the Leggett-Garg assumptions
of macro-realism. In order to demonstrate this result, we derive generalisations
of the Leggett-Garg inequalities originally put forward by Jordan
et al. \cite{jordan_kickedqndlg2}.}

\textcolor{black}{In Section V of this paper, we consider the case
where, $k=2$, which corresponds to a Kerr nonlinearity. As explained
above, this Hamiltonian has been realised experimentally for Bose-Einstein
condensates \cite{collapse-revival-bec,wrigth-walls-gar} and, more
recently, using superconducting circuits \cite{collapse-revival-super-circuit}.
In both experiments, the collapse and revival of a coherent state
were observed, with intermediate states formed that are strongly suggestive
of cat-states. In Section IV of this paper, we demonstrate violations
of the Leggett-Garg inequality for even values of $k$ greater than
$2.$ This case is presented because of the simplicity of the predictions
for violating a Leggett-Garg inequality, and because larger violations
are predicted. Although more difficult with current technology, an
experiment with higher quantum nonlinearities $k>2$ may become feasible.
For instance, such nonlinearities have been proposed for the generation
of entangled triplets of photons \cite{triple-non}.}

\textcolor{black}{In summary, our proposal for testing Leggett-Garg
macro-realism with $k=2$ corresponds to the highly nonlinear regime
of the experiments of Greiner et al. \cite{collapse-revival-bec}
and Kirchmair et al. \cite{collapse-revival-super-circuit}. The
times we propose for the Leggett-Garg tests are within the timescale
over which these experiments demonstrate the collapse and revival
of the coherent state, suggesting a Leggett-Garg experiment to be
highly feasible. As summarised in the second paragraph of the Introduction
however, the macro-realism assumptions introduced by Leggett and Garg
involve an additional assumption about measurements. This can create
extra complexities for the experimental realisations of the Leggett-Garg
inequality. In Section VI we give a specific discussion of how one
may achieve the test of Leggett-Garg inequalities in the experiments
of Greiner et al. \cite{collapse-revival-bec} and Kirchmair et al.
\cite{collapse-revival-super-circuit}, based on the additional assumption
of }\textcolor{black}{\emph{stationarity}}\textcolor{black}{{} \cite{stat-1zhou,stat-2-neutrino}.
This assumption has been applied to demonstrate quantum coherence
and violations of Leggett-Garg inequalities for photons and neutrinos
\cite{stat-1zhou,stat-2-neutrino}. We also outline alternative strategies
based on weak measurements \cite{lgexpphotonweak,weak,weakLGbellexp_review,laura-lg-two-well}.
Despite the additional assumptions, we argue that a successful demonstration
of the violation of the Leggett-Garg inequality would give a rigorous
confirmation of the formation of the superposition cat-state at the
intermediate times. This is because the violation could not be achieved
for a mixture of coherent states. A discussion of the implications
and potential loopholes of such experiments is given in the Sections
VI and VII.}

\section{Generalized Leggett-Garg inequalities}

\textcolor{black}{In this Section, we derive the Leggett-Garg inequalities
to be used in the remaining sections of the paper. Leggett and Garg
introduced two premises as part of their definition of macroscopic
realism. The first premise is called ``macroscopic realism per se'':
a system must always be in one }\textcolor{black}{\emph{or}}\textcolor{black}{{}
other of the macroscopically distinguishable states, prior to any
measurement being made. The second premise is called ``macroscopic
noninvasive measurability'': a measurement exists that can reveal
which state the system is in, with a negligible effect on the subsequent
macroscopic dynamics of the system. These two premises are used to
derive a Leggett-Garg inequality.}

\textcolor{black}{Let us assume that at each time $t_{i}$, the system
is in one of two macroscopically distinct states symbolised by $\varphi_{1}(t_{i})$
and $\varphi_{2}(t_{i})$. At each of three times $t_{1}$, $t_{2}$,
$t_{3}$, a measurement $\hat{S}(t_{i})$ is performed to indicate
which state the system is in. In the original Leggett-Garg treatment,
the result of the measurement is denoted by $S(t_{i})=1$ if the system
is found to be in $\varphi_{1}(t_{i})$, and $S(t_{i})=-1$ if the
system is found to be in $\varphi_{2}(t_{i})$. This choice was in
analogy with the Pauli spin-$1/2$ outcomes chosen by Bell in his
derivation of the famous Bell inequalities \cite{bell-1}.}

\textcolor{black}{To create greater flexibility, we will allow results
of the measurements to be denoted by a value $S(t_{i})$ which has
a magnitude less than $1$. In particular, this will allow us to define
an outcome to be $0$. A similar approach was taken for Bell inequalities
and Clauser-Horne inequalities, when generalised to account for outcomes
of no detection of a particle \cite{bell-review}. We will also allow
for the possibility that the macroscopically distinct states $\varphi_{1}(t_{i})$
and $\varphi_{2}(t_{i})$ defined at the different times $t_{i}$
can be different. This proves useful in deriving inequalities that
can be violated for the dynamics under the Hamiltonian (\ref{eq:hnl}).}

\textcolor{black}{Let us therefore consider a general case, where
the outcome of the measurement $\hat{S}(t_{i})$ is denoted by $S(t_{i})=x_{1}(t_{i})$
if the system is found to be in $\varphi_{1}(t_{i})$, and denoted
by $S(t_{i})=x_{2}(t_{i})$ if the system is found to be in $\varphi_{2}^{(i)}$,
where $|x_{1,2}^{(i)}|\leq1$. For the initial time $t_{1}$, we will
take $x_{1}(t_{1})=+1$ and $x_{2}(t_{1})=-1$, as in the original
derivation of Leggett and Garg \cite{legggarg}. However, at the times
$t_{2}$ and $t_{3}$, the values can be less than $1$. We will extend
the derivation of the Leggett-Garg inequality derived by Jordan et
al \cite{jordan_kickedqndlg2} to account for this case.}

\textcolor{black}{Following the original derivation given by Leggett
and Garg \cite{legggarg}, assuming macroscopic realism per se, we
can assign to the system at the times $t_{i}$ a hidden variable $\lambda_{i}$,
that }\textcolor{black}{\emph{predetermines}}\textcolor{black}{{} the
value of $S(t_{i})$ prior to the measurement $\hat{S}(t_{i})$. According
to the assumption of macroscopic realism per se, the system is always
in one }\textcolor{black}{\emph{or}}\textcolor{black}{{} other of the
states. Hence, the value of the hidden variable is a predetermined
property of the system, regardless of whether the measurement $\hat{S}(t_{i})$
takes place. Considering the three different times $t_{i}$, we consider
three hidden variables $\lambda_{1}$, $\lambda_{2}$ and $\lambda_{3}$.
Assuming macroscopic realism, the value $S(t_{i})$ of the measurement
is determined by the value of the hidden variable $\lambda_{i}$.
If the system at time $t_{i}$ is in state $\varphi_{1}(t_{i})$,
then we assign the value $x_{1}(t_{i})$ to the hidden variable $\lambda_{i}$.
If the system at time $t_{i}$ is in state $\varphi_{2}(t_{i})$,
then we assign the value $x_{2}(t_{i})$ to the hidden variable $\lambda_{i}$.
The hidden variables $\lambda_{i}$ assume a value that coincides
with the values of the possible results $S(t_{i})$ of the measurement.
In the original Leggett-Garg analysis, the hidden variables therefore
assume a value of $+1$ or $-1$. In our generalised case, the hidden
variables assume values bounded by $1$ i.e. $|\lambda_{i}|\leq1$. }

\textcolor{black}{Always then, $|\lambda_{1}|=1$ and $|\lambda_{2}|,|\lambda_{3}|\leq1$.
This allows us to carry out the proof. Simple algebra shows that \cite{jordan_kickedqndlg2,legggarg}
\begin{equation}
\lambda_{1}\lambda_{2}+\lambda_{2}\lambda_{3}-\lambda_{1}\lambda_{3}\leq1\label{eq:alg}
\end{equation}
because each $\lambda_{i}$ is bounded by $1$. This may be proved
straightforwardly. The value of $\lambda_{1}$ is either $1$ or $-1$.
Suppose $\lambda_{1}=1$. The maximum value of the function $F=\lambda_{2}+\lambda_{2}\lambda_{3}-\lambda_{3}$
over the domain $|\lambda_{2}|,|\lambda_{3}|\leq1$ is readily determined
to be $1$. This can be seen by graphical means. Alternatively, this
can be seen by noting the stationary point is given by coordinates
$(\lambda_{2},\lambda_{3})=(1,-1)$ and by considering the values
of $F$ at the boundaries: When $\lambda_{2}=1$, $F=1$; when $\lambda_{2}=-1$,
$F=-1-2\lambda_{3}\leq1$; when $\lambda_{3}=1$, $F=2\lambda_{2}-1\leq1$;
when $\lambda_{3}=-1$, $F=1$. Thus, $F\leq1$ for all $\lambda_{2}$,
$\lambda_{3}$ in the domain $|\lambda_{2}|,|\lambda_{3}|\leq1$.
Next, one considers $\lambda_{1}=-1$, to show in this case it is
also true that $F\leq1$.}

\textcolor{black}{Following the original derivation of the Leggett-Garg
inequality, one now applies the second assumption of macro-realism.
One assumes that a macroscopically noninvasive measurement is made
on the system to determine the value $S(t_{i})$ at each time $t_{i}$.
Then one considers the two-time correlation functions defined by $\langle S(t_{i})S(t_{j})\rangle$.
Using the premises, these moments are given $\langle S(t_{i})S(t_{j})\rangle=\langle\lambda_{i}\lambda_{j}\rangle$,
this leads to the Leggett-Garg inequality 
\begin{equation}
{\color{black}{\color{black}\langle S_{1}S_{2}\rangle+\langle S_{2}S_{3}\rangle-\langle S_{1}S_{3}\rangle}\leq1}\label{eq:lginequality}
\end{equation}
where for simplicity of notation we have introduced the abbreviation
$S_{i}\equiv S(t_{i})$ and $\langle S_{i}S_{j}\rangle\equiv\langle S(t_{i})S(t_{j})\rangle$.
The inequality is derived based on the assumptions of macro-realism.
The violation of the inequality for an appropriate experiment therefore
falsifies the macro-realism premises. This inequality was originally
derived in Ref. \cite{jordan_kickedqndlg2} for the case where $\lambda_{i}=\pm1$.
Violations of the inequality are predicted for states evolving according
to $\langle S_{i}S_{j}\rangle=\cos2(t_{j}-t_{i})$, as can be seen
by putting}\foreignlanguage{australian}{\textcolor{black}{{} $t_{1}=0$,
$t_{2}=\pi/6$, $t_{3}=\pi/3$ (or $t_{3}=5\pi/12$)}}\textcolor{black}{.}

\textcolor{black}{The obvious difficulty with carrying out a Leggett-Garg
experiment is the evaluation of the moments $\langle S_{i}S_{j}\rangle$
which are made under the assumption of a macroscopically noninvasive
measurement. There are several ways the moment $\langle S_{2}S_{3}\rangle$
can be evaluated for an experimental test of the inequality (refer
Refs. \cite{experiment-lg,jordan_kickedqndlg2,lgexpphotonweak,emeryreview,Mitchell,NSTmunro,laura-lg-two-well,massiveosci,small-systemlg,stat-1zhou,stat-2-neutrino}).
Experiments require justification that any measurement made at time
$t_{2}$ does not interfere with the subsequent macroscopic evolution
of the system. One method proposed in the original Leggett and Garg
paper is an ideal negative-result measurement. Another approach is
to make a weak measurement of the type proposed by Aharonov, Albert
and Vaidman \cite{weak,weakLGbellexp_review,lgexpphotonweak,laura-lg-two-well,weak-noon}.
Such a weak measurement does not fully collapse the wave function
at $t_{2}$, but allows one to infer the average $\langle S_{2}S_{3}\rangle$
over a series of runs. Alternatively, one may argue along the lines
of ``measure and re-prepare'' and ``stationarity'' \cite{stat-1zhou,stat-2-neutrino}.
This allows a test of the inequality by measuring two-time ensemble
averages only. The argument is as follows: If the system is indeed
in one of the states $\psi_{1}(t_{2})$ and $\psi_{2}(t_{2})$ at
time $t_{2}$, the experimentalist can determine $\langle S_{2}S_{3}\rangle$
by first measuring which of the states the system is in at $t_{2}$,
and then re-preparing that state (either $\psi_{1}(t_{2})$ and $\psi_{2}(t_{2})$),
to determine the value of $S_{3}$ at the later time $t_{3}$. The
details of this approach will be given in Sections IV and VI.}

\section{Model}

\textcolor{black}{In this paper, we show how the Leggett-Garg inequalities
can be violated for dynamical cat-states formed under the evolution
of a nonlinear interaction. In this Section, we explain the theoretical
predictions for the dynamical solutions. Following Yurke and Stoler
\cite{yurke-stoler}, we consider the evolution of a single mode system
prepared in a coherent state under the influence of a nonlinear Hamiltonian
written in the Schrodinger picture as}

\textcolor{black}{
\begin{equation}
H=\omega\hat{n}+\Omega\hat{n}^{k}\label{eq:ham}
\end{equation}
We will restrict to consider $k$ even. The odd case requires a different
analysis, because the evolution is different and gives rise to different
types of cat-states. The anharmonic term is proportional to $\hat{n}^{k}$
where $\hat{n}$ is the mode number operator and the integer $k>1$
represents the order of the nonlinearity. The $\omega$ is the frequency
of the harmonic oscillator and we choose units such that $\hbar=1$.
In the interaction picture, the evolution of the state can be readily
determined. The initial coherent state is of the form}

\textcolor{black}{
\begin{equation}
\left|\alpha\right\rangle =\exp[-\frac{\left|\alpha\right|^{2}}{2}]\underset{n}{\sum}\alpha^{n}\frac{1}{\sqrt{n!}}\left|n\right\rangle 
\end{equation}
where $\left|n\right\rangle $ is the $n$-particle eigenstate. The
state after a time $t$ is }

\textcolor{black}{
\begin{equation}
\left|\alpha,t\right\rangle =\exp[-\frac{\left|\alpha\right|^{2}}{2}]\underset{n}{\sum}\alpha^{n}\frac{\exp(-i\phi_{n})}{\sqrt{n!}}\left|n\right\rangle 
\end{equation}
where $\phi_{n}=\varOmega tn^{k}$. }

\textcolor{black}{It is known that at certain times the system evolves
into a superposition of distinct coherent states \cite{milburn-holmes,wrigth-walls-gar,yurke-stoler}.
For $k>2$, after a time $t=\pi/2\Omega$ the system is in a cat-state
with coherent amplitudes $\pi$ out of phase. At $t=\pi/\Omega$,
the system is again in a coherent state $|-\alpha\rangle$. At double
this time, there is a revival back to the original coherent state
$|\alpha\rangle$. Thus we observe cyclic behaviour \cite{yurke-stoler,wrigth-walls-gar,collapse-revival-bec,collapse-revival-super-circuit}
and the sign of $\Omega$ acts only to reverse the direction of evolution
of the states. Plots of the $Q$ functions representing the different
states are illustrated in Figure 1 for even values of $k$ greater
than $2$.}
\begin{figure}[b]
\textcolor{black}{\includegraphics[width=0.5\columnwidth]{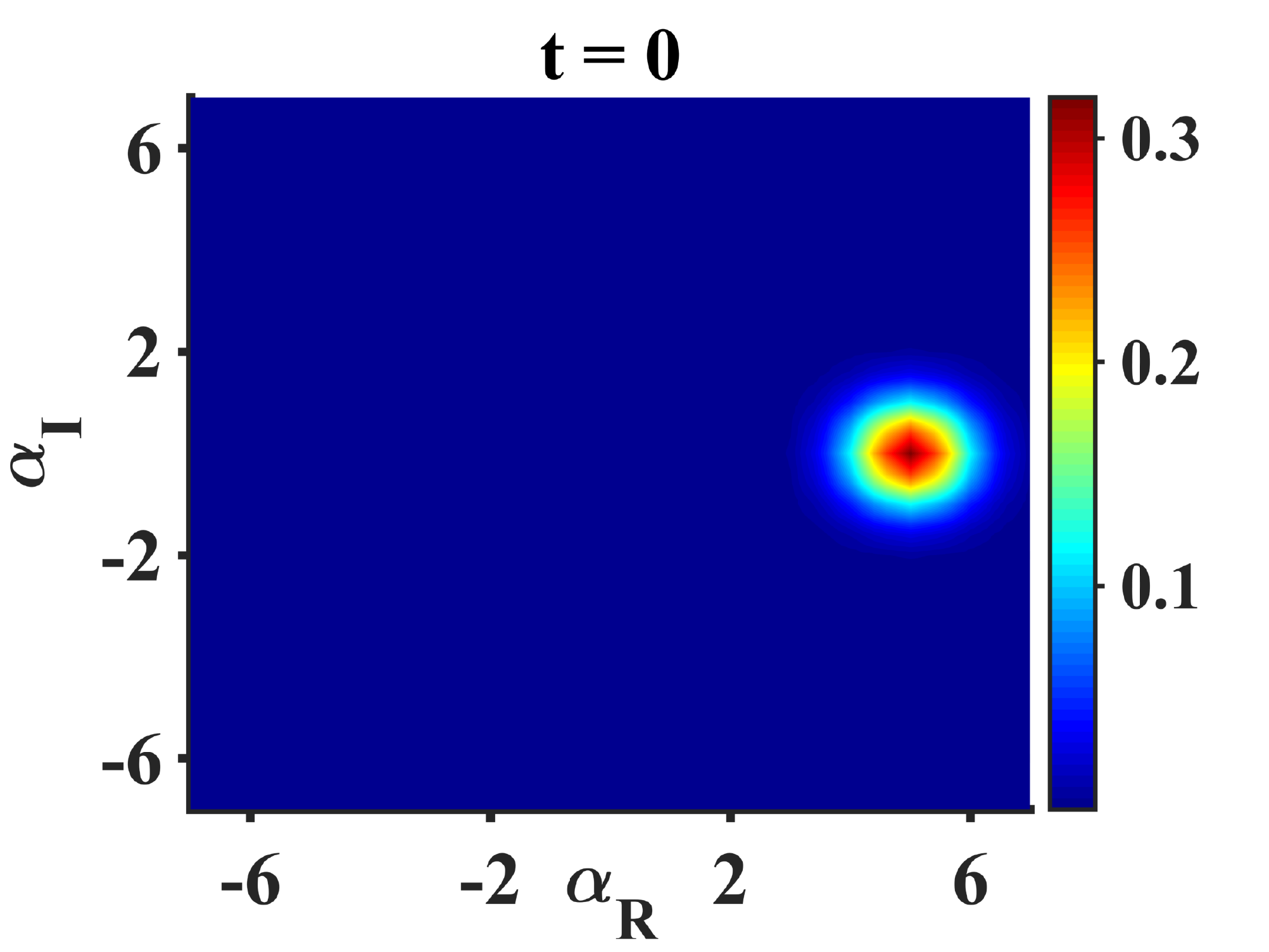}\includegraphics[width=0.5\columnwidth]{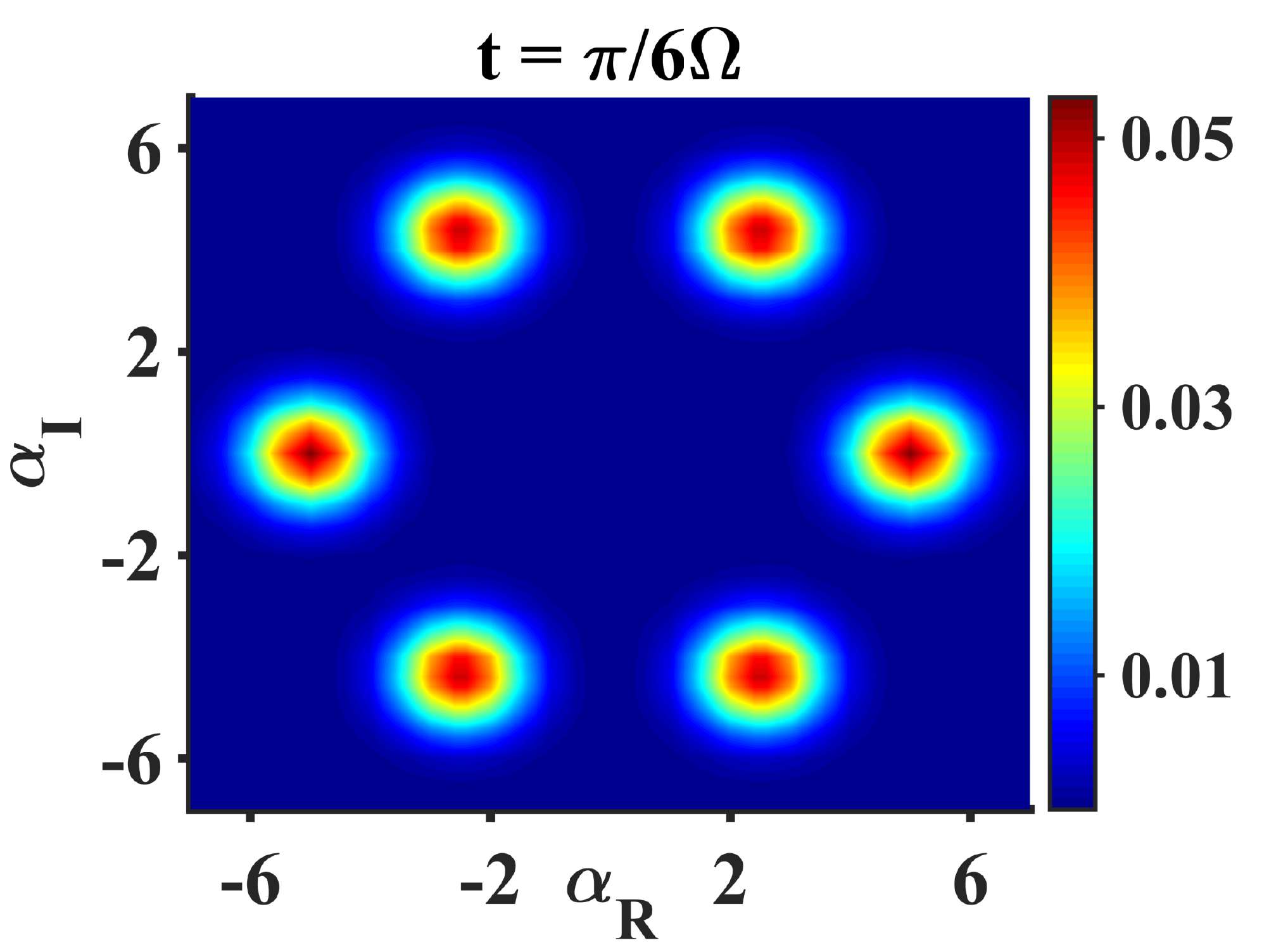}}

\textcolor{black}{\includegraphics[width=0.5\columnwidth]{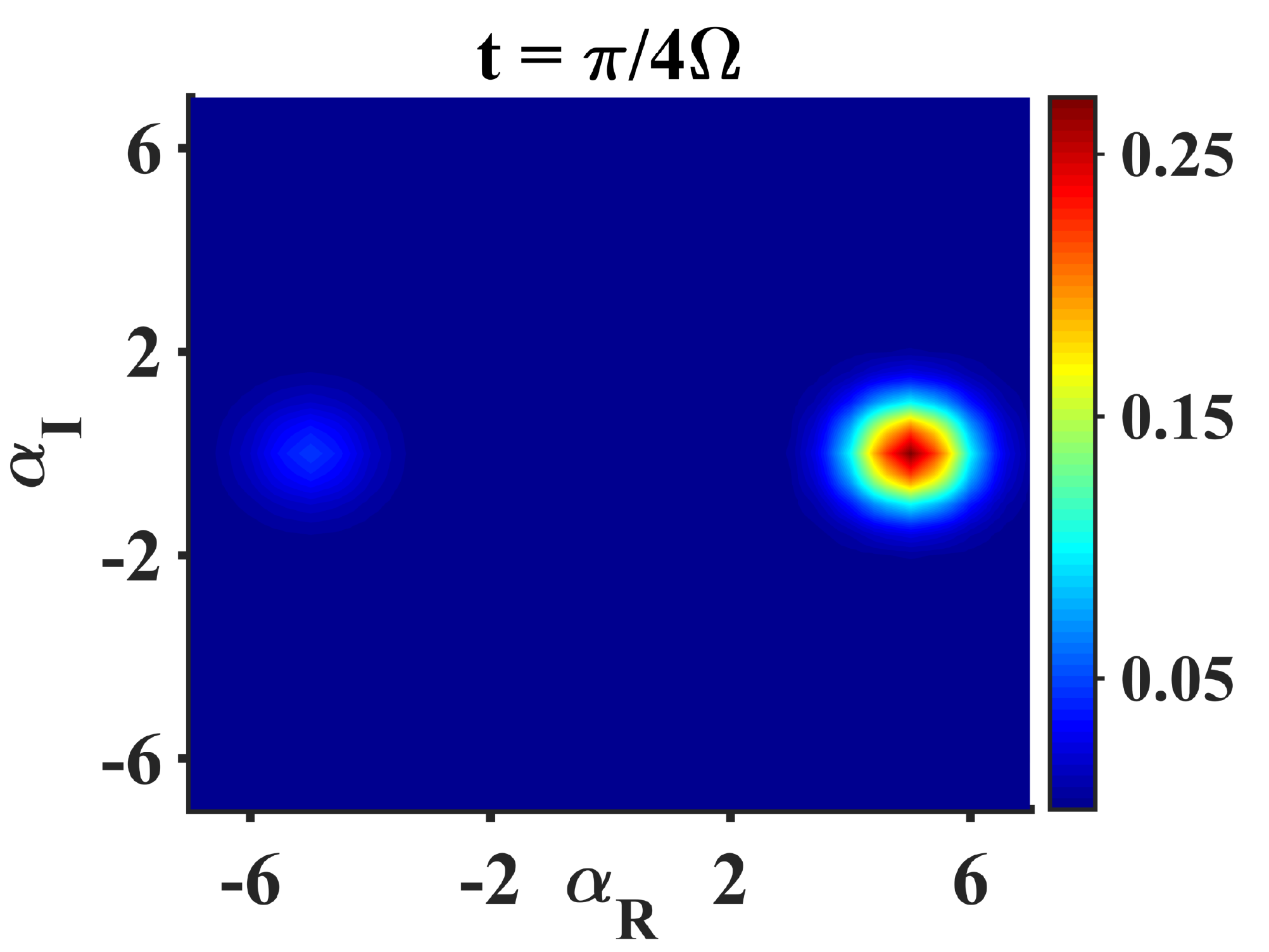}\includegraphics[width=0.5\columnwidth]{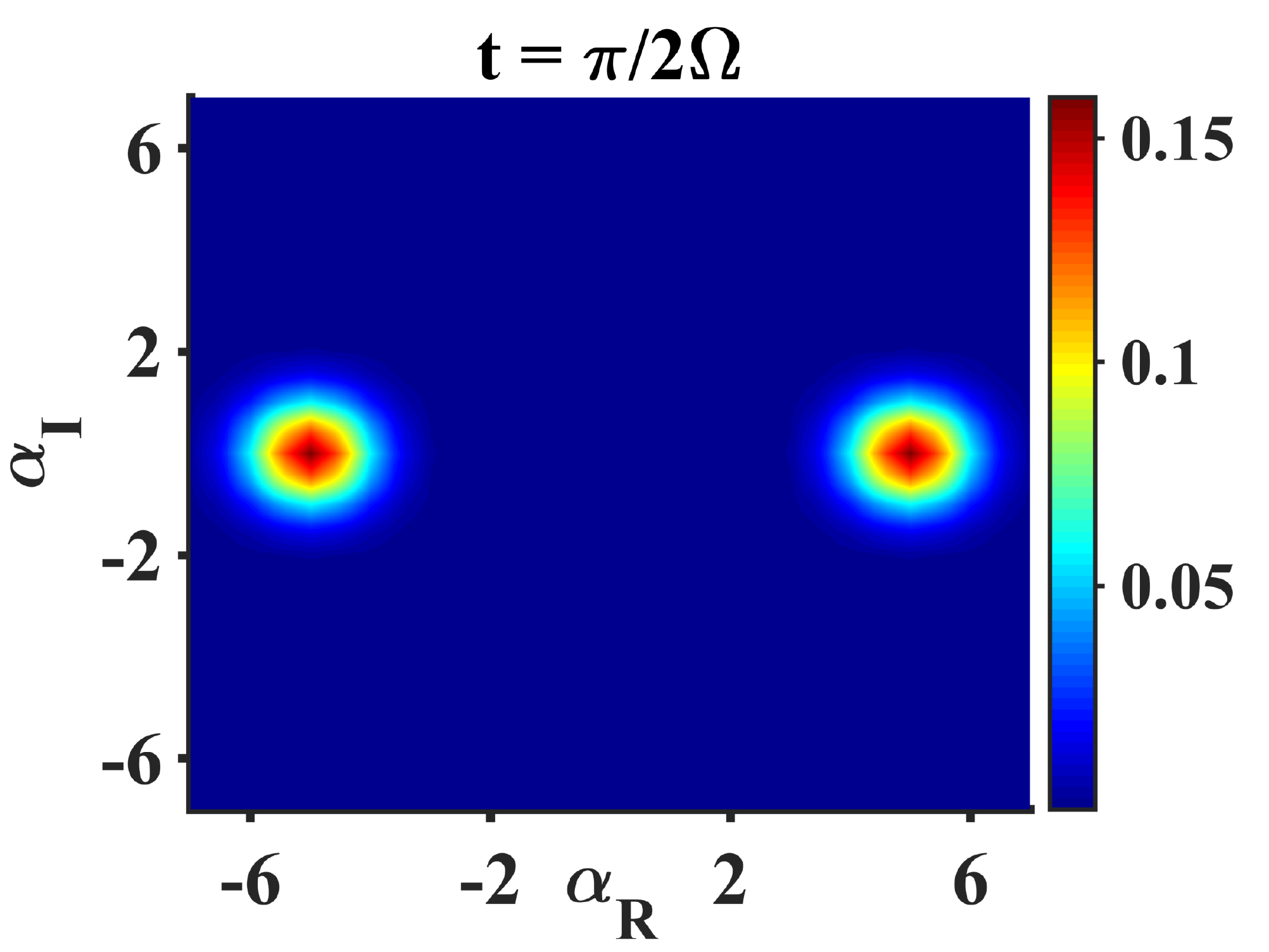}}

\textcolor{black}{\includegraphics[width=0.5\columnwidth]{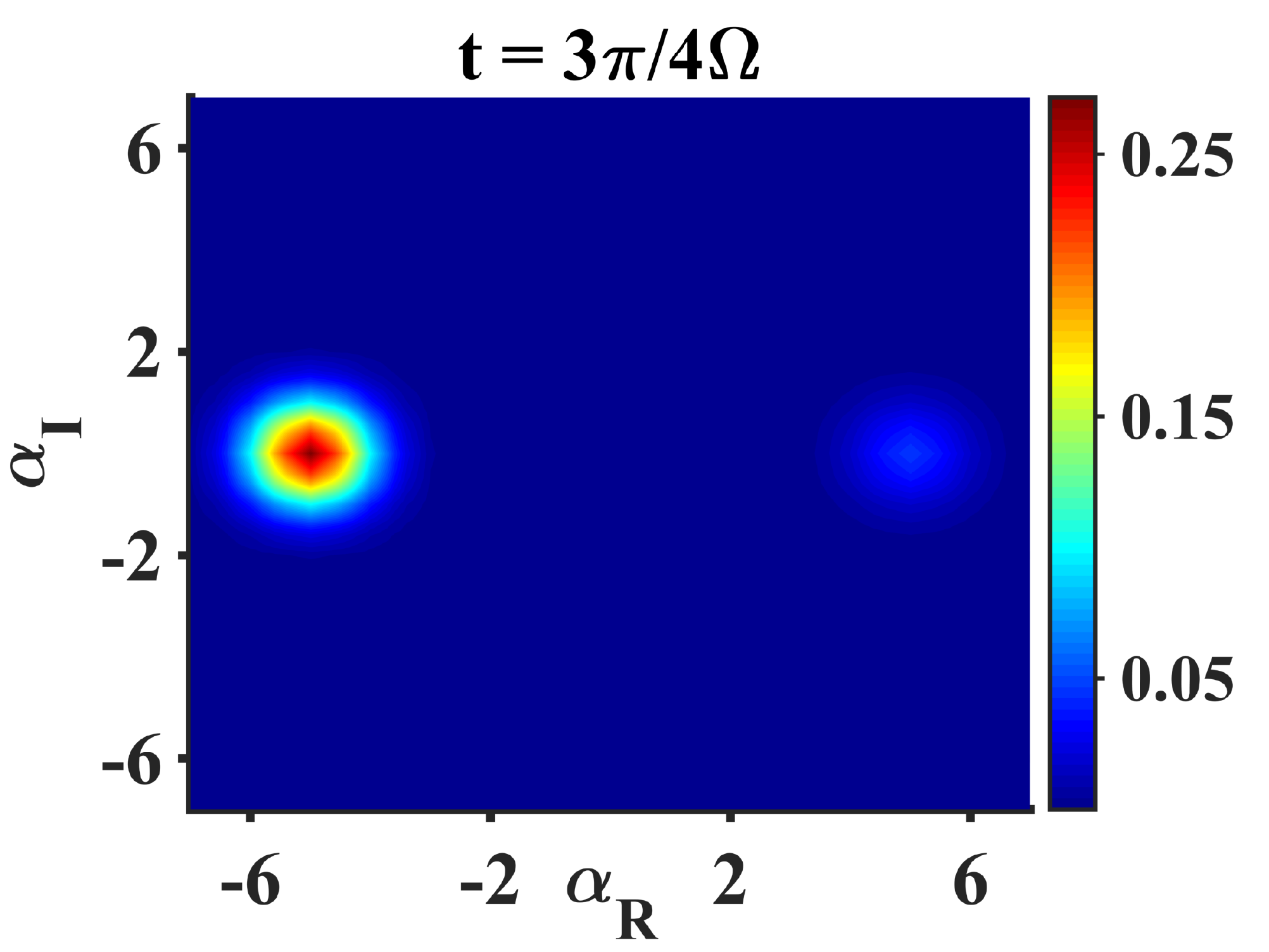}\includegraphics[width=0.5\columnwidth]{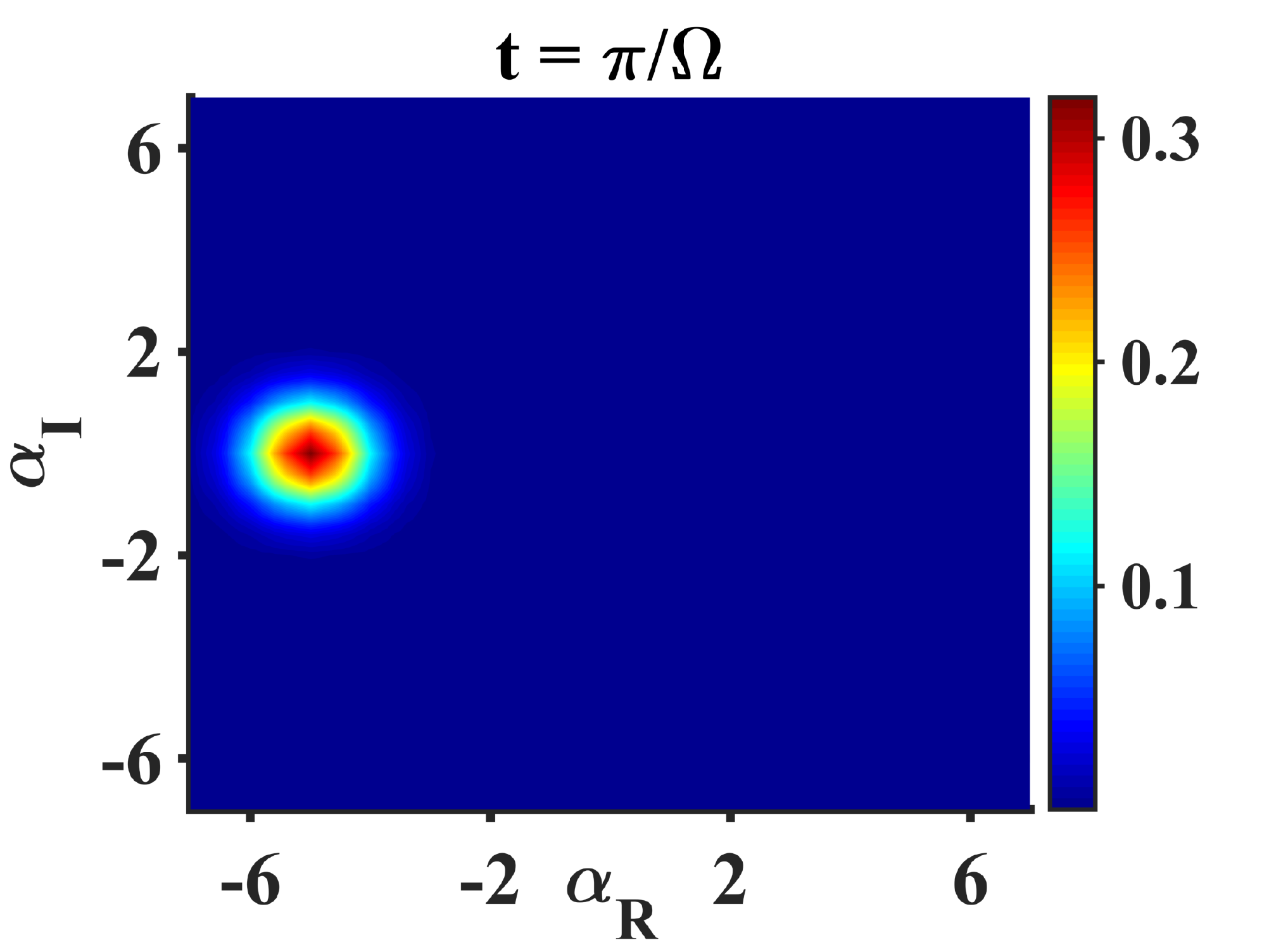}}

\textcolor{black}{\caption{\textcolor{black}{The evolution of multi-component cat-states for
a system described by the nonlinear Hamiltonian eq. (\ref{eq:ham})
where $k$ is any even number greater than $2$. The c}ontour graphs
show surface plots of $Q(\alpha)$ \textcolor{black}{with $\alpha_{0}=5$.
Plots are given for the times $t=0$, $t=\pi/6\Omega$, $t=\pi/4\Omega$,
$t=\pi/2\Omega$, $t=3\pi/4\Omega$ and $t=\pi/\Omega$.}\textcolor{red}{}\textcolor{blue}{}}
}
\end{figure}

\textcolor{black}{For the purpose of the Leggett-Garg tests, the times
$t_{2}$ and $t_{3}$ will correspond to the system being in a cat-superposition
state of some sort, where the amplitudes of the coherent states are
well separated in phase space. For instance, Figure 1 shows such cat-superposition
states at times $t=\pi/6\Omega$, $t=\pi/4\Omega$, $t=\pi/2\Omega$
and $t=3\pi/4\Omega$. We allow in general that the cat-states may
be different at the different times. This deviates from the traditional
Leggett-Garg test, where the system is in a superposition of the same
two states at all times. The generalised Leggett-Garg inequalities,
described in Section II, allow more flexibility to analyst Leggett-Garg
violations.}

\section{Leggett-Garg violations for nonlinearity $k>2$, $k$ even}

\textcolor{black}{In this Section, we consider $k>2$ and $k$ even.
We take $t_{1}=0$, when the system is prepared in a coherent state
$|\alpha\rangle$ (where $\alpha$ is real) and consider the subsequent
times $t_{2}=\pi/4\Omega$, $t_{3}=3\pi/4\Omega$. The analytic expression
at time $t=\pi/4\Omega$ can be readily evaluated \cite{yurke-stoler}.
When $t=\pi/4\Omega$, $e^{-i\Phi_{n}}=\exp(-i\pi n^{k}/4)$. Therefore
at $t=\pi/4\varOmega$ one has $e^{-i\varPhi_{n}}=(-1)^{n/2}$ when
$n$ is even and $k=2$, and $e^{-i\varPhi_{n}}=1$ when $n$ is even
and $k>2$. When $n$ is odd and $k$ even, one has $e^{-i\varPhi_{n}}=\frac{\sqrt{2}}{2}-i\frac{\sqrt{2}}{2}$
. }

\textcolor{black}{For our case of interest, when $k>2$ and $k$ is
even, the state generated at time $t=\pi/4\Omega$ is}

\textcolor{black}{
\begin{equation}
\left|\alpha,\pi/4\varOmega\right\rangle =\frac{1}{2}\{(1+e^{-i\pi/4})|\alpha\rangle+(1-e^{-i\pi/4})|-\alpha\rangle\}\label{eq:cat1}
\end{equation}
At $t=3\pi/4\Omega$, the solution is}

\textcolor{black}{
\begin{equation}
\left|\alpha,3\pi/4\varOmega\right\rangle =\frac{1}{2}\{(1-e^{-i\pi/4})|\alpha\rangle+(1+e^{-i\pi/4})|-\alpha\rangle\}
\end{equation}
This compares with 
\begin{equation}
\begin{array}{c}
\frac{1}{\sqrt{2}}(e^{-i\frac{\pi}{4}}|\alpha\rangle+e^{+i\frac{\pi}{4}}|-\alpha\rangle)\end{array}\label{eq:2cat-pi}
\end{equation}
at the time $t=\pi/2\Omega$. The $Q$ functions for the states generated
at the four times $t=0$, $\pi/4\Omega,$ $\pi/2\Omega$ and $\text{3\ensuremath{\pi}/4\ensuremath{\Omega}\ }$
are plotted in Figure 2. Also plotted in Figure 3 is the value of
the probability density $P(x)$ for a measurement $x$ on each of
the states, where $\hat{x}={\color{red}{\color{blue}{\color{black}\frac{1}{\sqrt{2}}}}}(\hat{a}+\hat{a}^{\dagger})$.
The calculations for $P(x)$ and the $Q$ function are outlined in
the Appendix.}
\begin{figure}[h]
\textcolor{black}{\includegraphics[width=0.5\columnwidth]{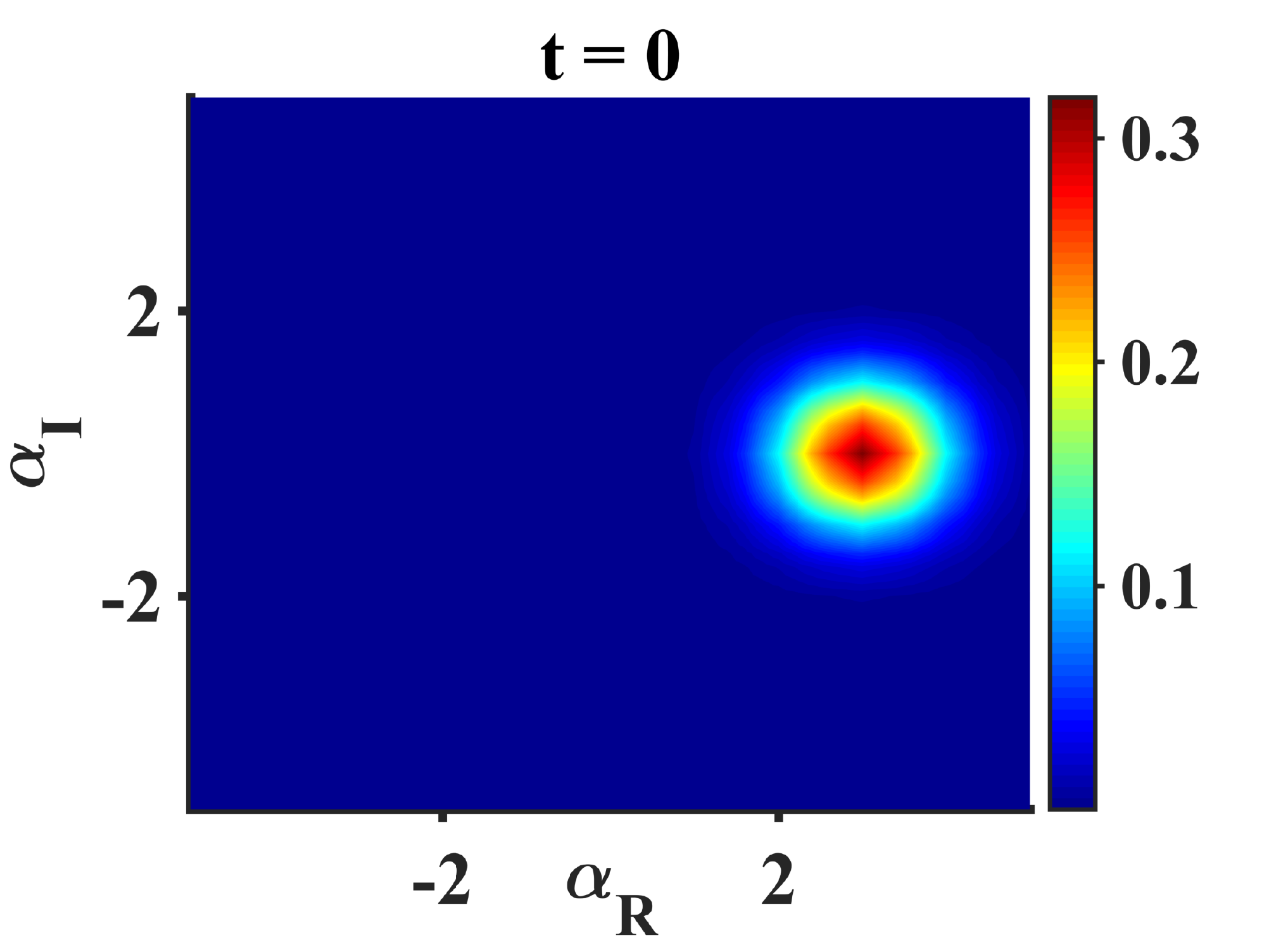}\includegraphics[width=0.5\columnwidth]{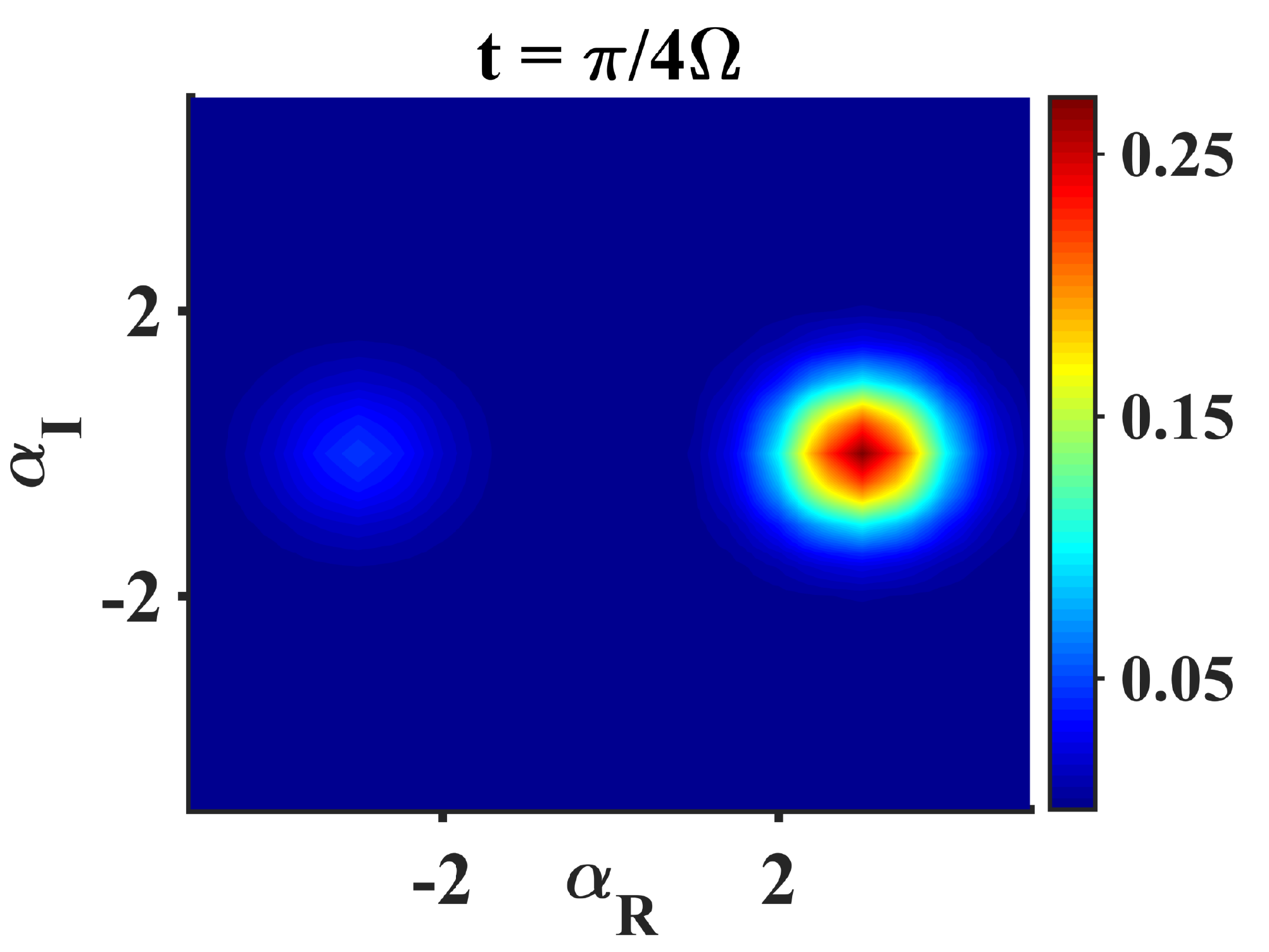}}

\textcolor{black}{\includegraphics[width=0.5\columnwidth]{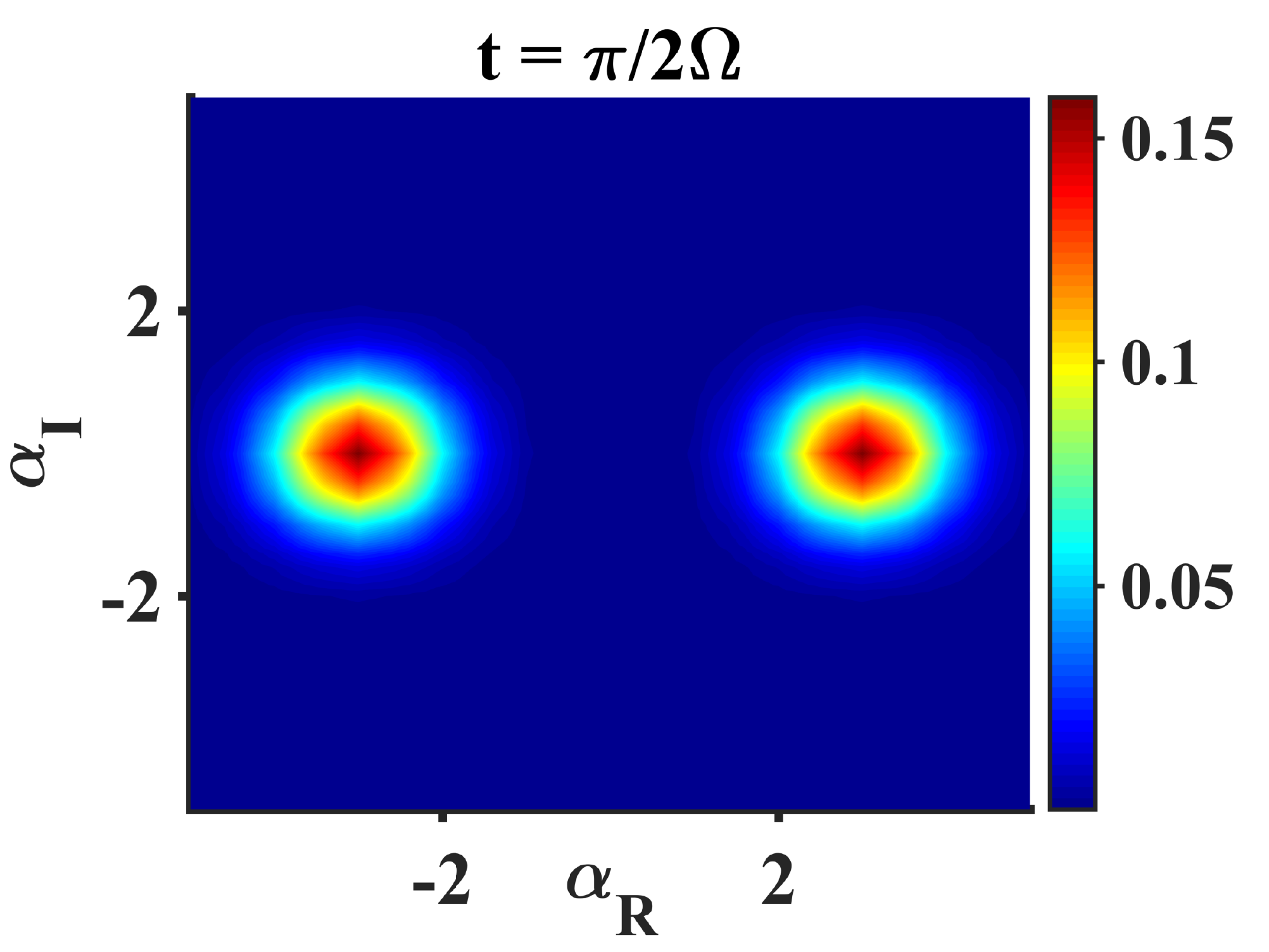}\includegraphics[width=0.5\columnwidth]{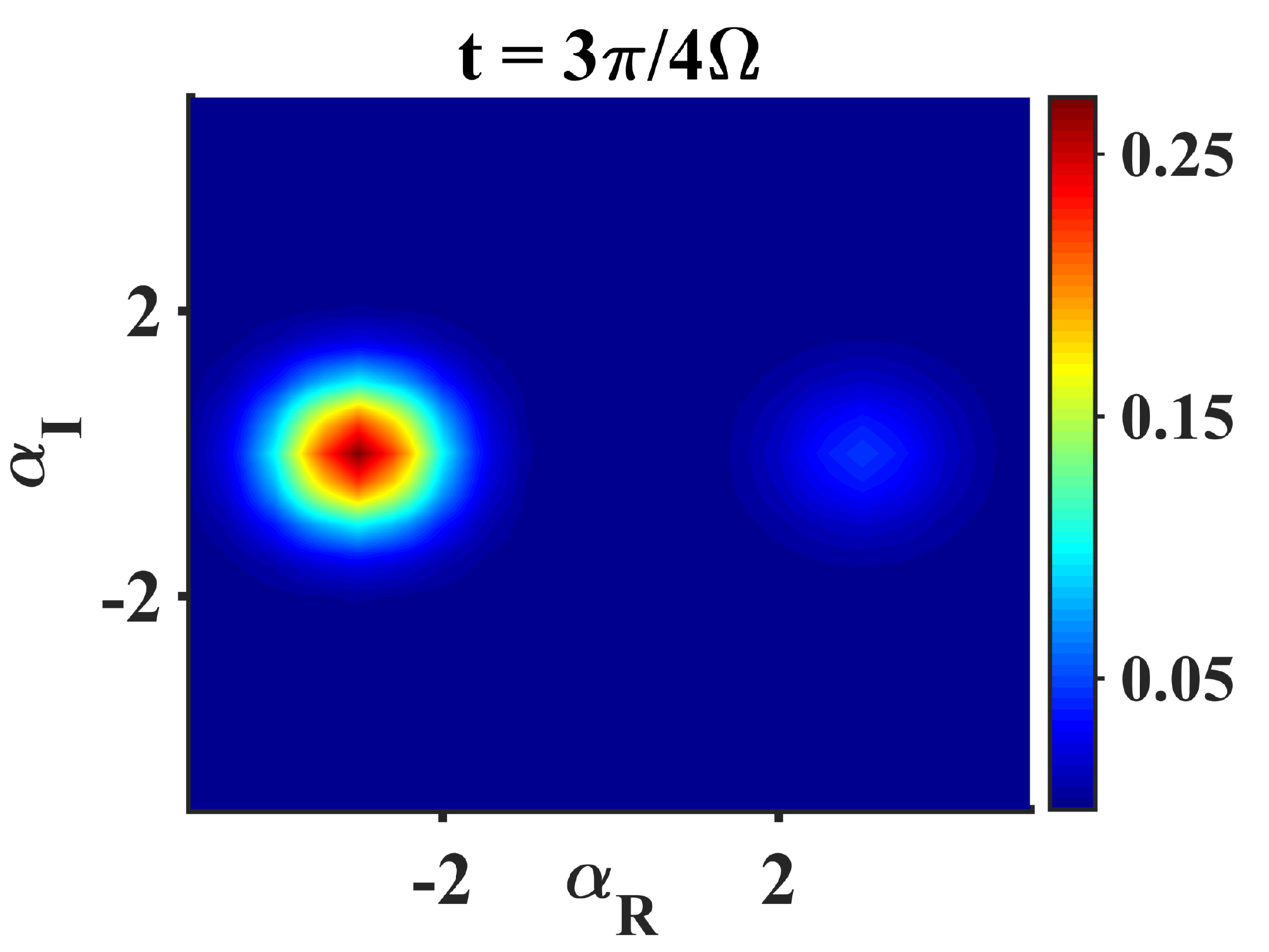}}

\textcolor{black}{\caption{\textcolor{black}{The evolution of two-component cat-states for a
system described by the nonlinear Hamiltonian eq. (\ref{eq:ham})
where $k$ is any even number greater than $2$. The con}tour graphs
show surface plots of $Q(\alpha)$ \textcolor{black}{with $\alpha_{0}=3$.
The plots are given for the times $t=0$, $t=\pi/4\Omega$, $t=\pi/2\Omega$
and $t=3\pi/4\Omega$.}}
}
\end{figure}

\textcolor{black}{}
\begin{figure}[h]
\begin{raggedright}
\textcolor{black}{\includegraphics[width=0.5\columnwidth]{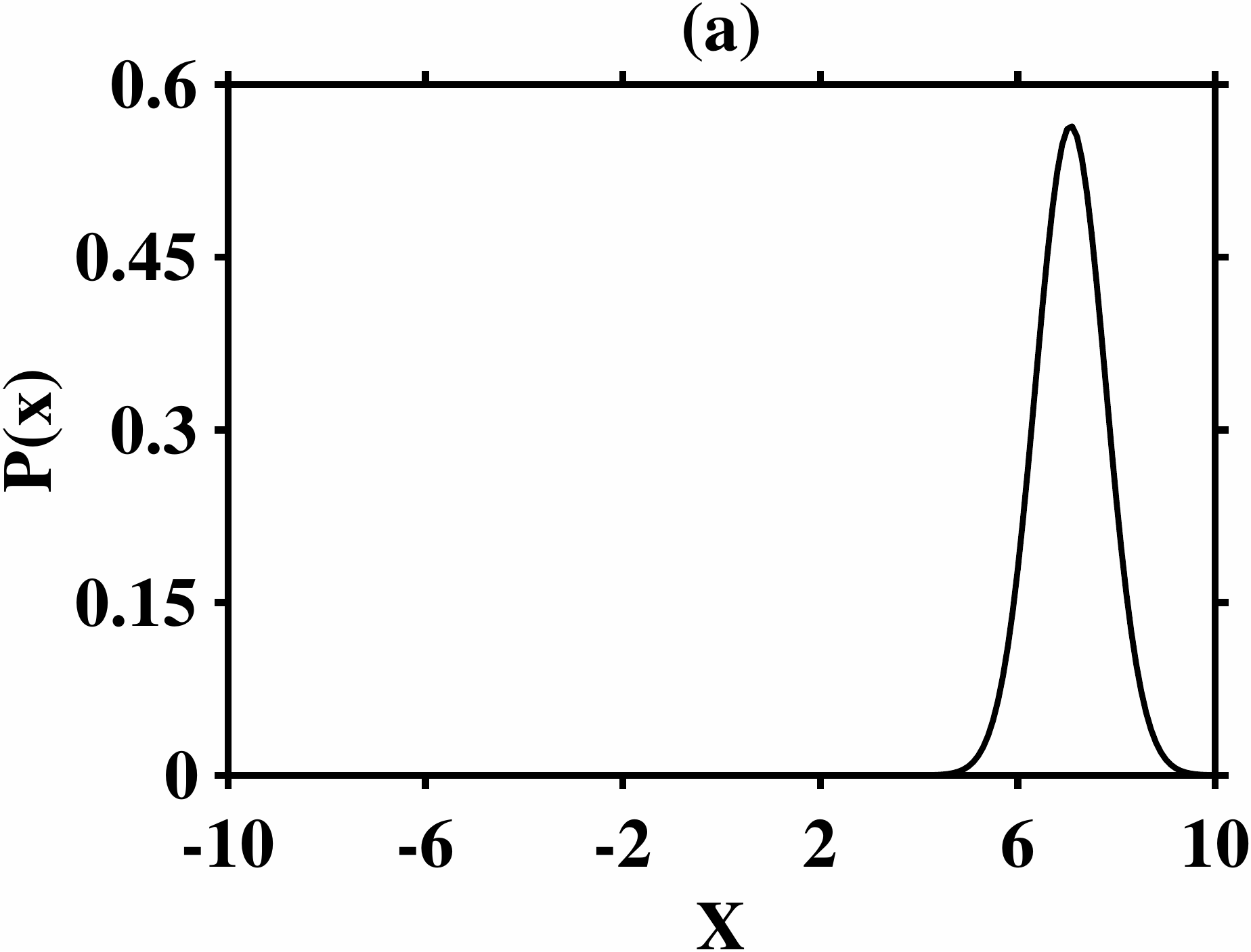}\includegraphics[width=0.5\columnwidth]{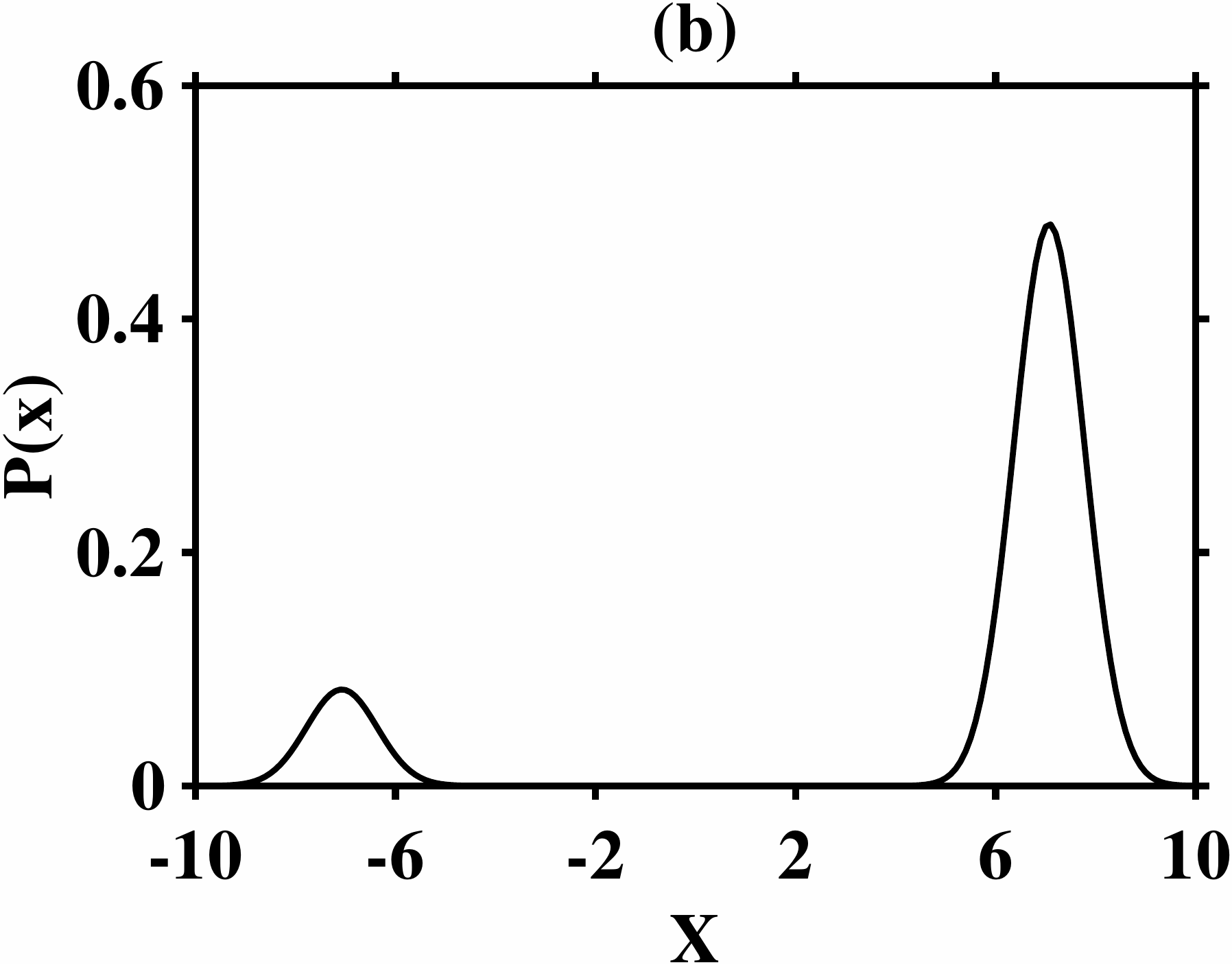}}
\par\end{raggedright}
\textcolor{black}{\includegraphics[width=0.5\columnwidth]{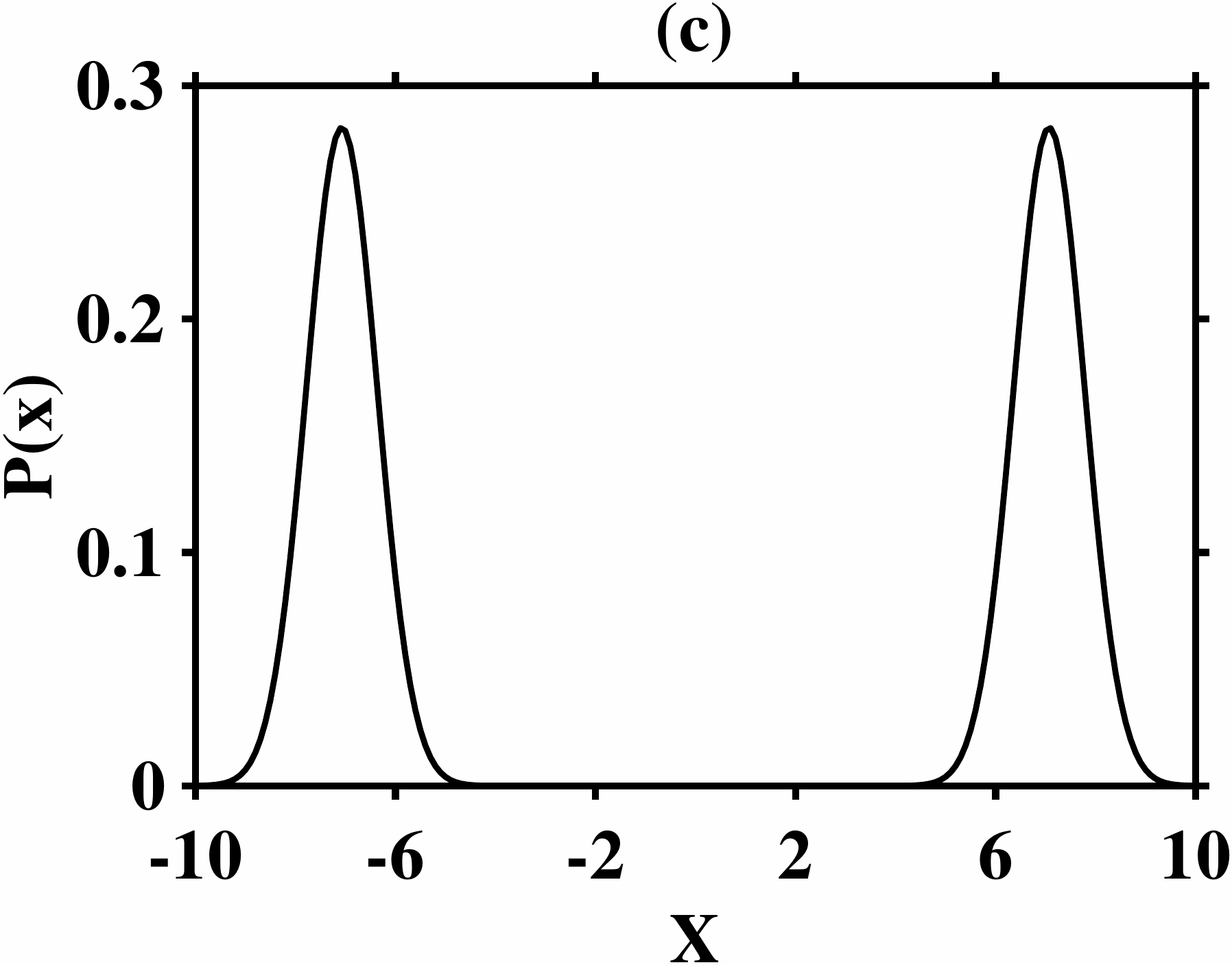}\includegraphics[width=0.5\columnwidth]{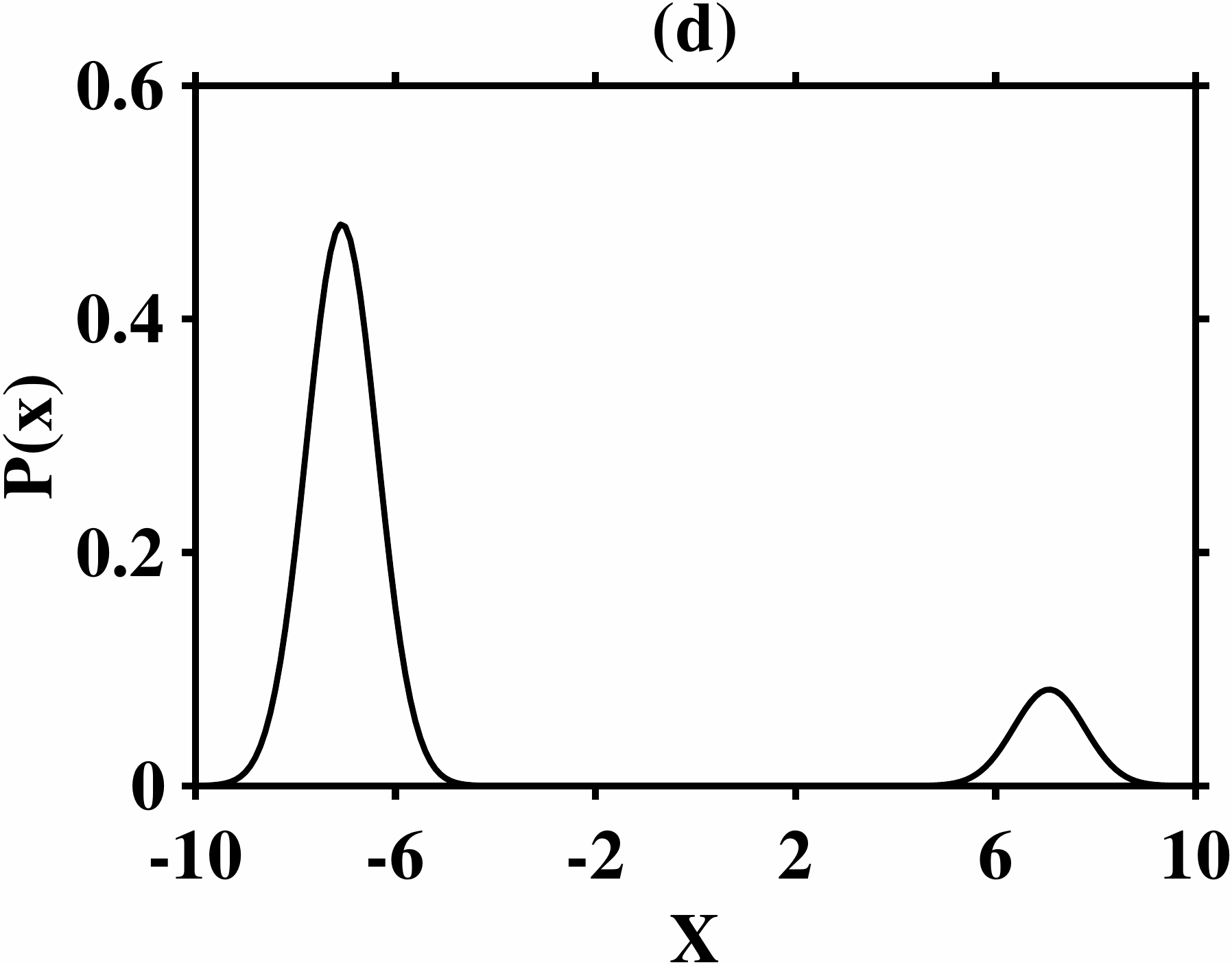}}

\textcolor{black}{\caption{\textcolor{black}{The evolution of two-component cat-states for a
system described by the nonlinear Hamiltonian eq. (\ref{eq:ham})
where $k$ is any even number greater than $2$. T}he graphs show
plots of $P(x)$ \textcolor{black}{with $\alpha=5$. The figures (a)-(d)
give $P(x)$ at $t=0$, $t=\pi/4\Omega$, $t=\pi/2\Omega$, and $t=3\pi/4\Omega$}
respectively.\label{fig:Data-for-EPR}\textcolor{black}{}}
}
\end{figure}

\textcolor{black}{We now evaluate the Leggett-Garg inequality (\ref{eq:lginequality})
using $t_{1}=0$, $t_{2}=\pi/4\Omega$ and $t_{3}=3\pi/4\Omega$.
Since $S_{1}=1$, we evaluate $\langle S_{1}S_{2}\rangle$ as $\langle S_{2}\rangle=P_{+}^{(2)}-P_{-}^{(2)}$,
where $P_{+}^{(i)}=\int_{0}^{\infty}P(x)dx$ is the probability of
result for $x$ being greater than or equal to $0$ at time $t_{i}$,
and $P_{-}^{(i)}=\int_{-\infty}^{0}P(x)dx$ is the probability of
result for $x$ being less than $0$ at time $t_{i}$. By integration,
we find $P_{+}^{(2)}=0.8535$ and $P_{-}^{(2)}=0.1465$ ($\alpha\geq2$).
 Therefore $\langle S_{1}S_{2}\rangle=0.7070$. Similarly, we see
that $\langle S_{1}S_{3}\rangle=-\langle S_{1}S_{2}\rangle$ (refer
Figure 2). }

\subsection*{\textcolor{black}{Determining $\langle S_{2}S_{3}\rangle$ }}

\textcolor{black}{We have evaluated the two-time moments $\langle S_{1}S_{2}\rangle$
and $\langle S_{1}S_{3}\rangle$ straightforwardly without consideration
of any intermediate measurement that might be made at time $t_{2}$.
This is justified for two reasons. First, in the Leggett-Garg derivation,
such a measurement is required to be macroscopically noninvasive (refer
Section VI). Also, the moments $\langle S_{1}S_{2}\rangle$ and $\langle S_{1}S_{3}\rangle$
are predicted to be a function of the time }\textcolor{black}{\emph{differences}}\textcolor{black}{{}
$t_{2}-t_{1}$ and $t_{3}-t_{1}$ respectively, which justifies the
assumption of stationarity, that the two-time moments are invariant
under time translation. }

\textcolor{black}{These considerations justify an approach that can
be used to determine $\langle S_{2}S_{3}\rangle$. To evaluate $\langle S_{2}S_{3}\rangle$,
we use the ``measure and re-prepare'' approach, discussed at the
end of the Section II. Specifically, we will use the expression
\begin{equation}
\langle S_{2}S_{3}\rangle=P_{+}^{(2)}\langle S_{2}S_{3}\rangle_{+}+P_{-}^{(2)}\langle S_{2}S_{3}\rangle_{-}\label{eq:smix}
\end{equation}
where $P_{\pm}^{(2)}$ is the probability that the system at time
$t_{2}$ has a positive or negative value for $x$. This probability
can be measured experimentally. For sufficiently large $\alpha$,
this probability is equal to the probability the system can be found
to be in the $|\pm\alpha\rangle$ state. Here we denote $\langle S_{2}S_{3}\rangle_{+}$
as the average of $\langle S_{2}S_{3}\rangle$ given the state is
prepared in the state $|\alpha\rangle$ at time $t_{2}$. Similarly,
$\langle S_{2}S_{3}\rangle_{-}$ is the average of $S$ after a time
$t$ given the state is prepared in the state $|-\alpha\rangle$ at
time $t_{2}$. Recall from Section II that we define the two-time
correlation as $\langle S_{2}S_{3}\rangle\equiv\langle S(t_{2})S(t_{3})\rangle$.}

\textcolor{black}{The expression (\ref{eq:smix}) is justified if
we assume the Leggett-Garg premises. At time $t_{2}$, the system
is in a superposition of two states $|\psi_{1}\rangle$ and $|\psi_{2}\rangle$
\begin{equation}
|\psi\rangle=c_{-}|\psi_{1}\rangle+c_{+}|\psi_{2}\rangle\label{eq:sup-1}
\end{equation}
where $|\psi_{1}\rangle=|-\alpha\rangle$ and $|\psi_{2}\rangle=|\alpha\rangle$.
Here $c_{-}$ and $c_{+}$ are probability amplitudes, where $P_{\pm}^{(2)}=|c_{\pm}|^{2}$
for large $\alpha$. The assumption of Leggett-Garg macro-realism
is that the system is in }\textcolor{black}{\emph{one or the other}}\textcolor{black}{{}
of the states $|\psi_{1}\rangle$ and $|\psi_{2}\rangle$ at time
$t_{2}$ (with probabilities $P_{+}^{(2)}$ and $P_{-}^{(2)}$ respectively).
If we assume that at time $t_{2}$ the system }\textcolor{black}{\emph{was}}\textcolor{black}{{}
in state $|\alpha\rangle$, then this is the initial state for the
calculation of $\langle S_{2}S_{3}\rangle$, which is then represented
as $\langle S_{2}S_{3}\rangle_{+}$. We see that because $t_{3}-t_{2}=\pi/2\Omega$,
if the system is indeed in the state $|\alpha\rangle$ at time $t_{2}$,
then the system at the later time $t_{3}$ is in the symmetric state
with equal probability for $x>0$ and $x<0$, as evident from Figures
2 and 3. Therefore $\langle S_{2}S_{3}\rangle_{+}=0$. Similarly,
if we take that the system at time $t_{2}$ }\textcolor{black}{\emph{was}}\textcolor{black}{{}
in state $|-\alpha\rangle$, then we can show that $\langle S_{2}S_{3}\rangle_{-}=0$.
Therefore $\langle S_{2}S_{3}\rangle=0$. Thus, we evaluate the Leggett-Garg
term as
\begin{equation}
{\color{black}{\color{black}\langle S_{1}S_{2}\rangle+\langle S_{2}S_{3}\rangle-\langle S_{1}S_{3}\rangle}=1.414}
\end{equation}
}

\noindent \textcolor{black}{This shows a violation of the Leggett-Garg
inequality (\ref{eq:lginequality}). The term ``measure and re-prepare''
is used to describe this technique because in principle, one can measure
which state the system is in at time $t_{2}$, and then re-prepare
that state to determine $\langle S_{2}S_{3}\rangle$. The assumption
of stationarity however means that the moments $\langle S_{2}S_{3}\rangle_{\pm}$
(which are predicted to be dependent only on the time }\textcolor{black}{\emph{difference}}\textcolor{black}{{}
$t_{3}-t_{2}$) can be measured more conveniently in an independent
experiment at any later or prior time.}

\section{Leggett-Garg violations for nonlinearity $k=2$}

\textcolor{black}{For the case $k=2$, the evolution of the cat-states
is different to the case with $k\neq2$, $k$ even. We will consider
other time intervals in order to obtain a violation of the Leggett-Garg
inequality. The Appendix gives the analytical expressions for the
states and the probabilities $P(x)$ at different times of evolution.
The $Q$ functions for the states generated at a selection of different
times are plotted in Figures 4 and 5.}

\textcolor{black}{}
\begin{figure}[h]
\textcolor{black}{\includegraphics[width=1\columnwidth]{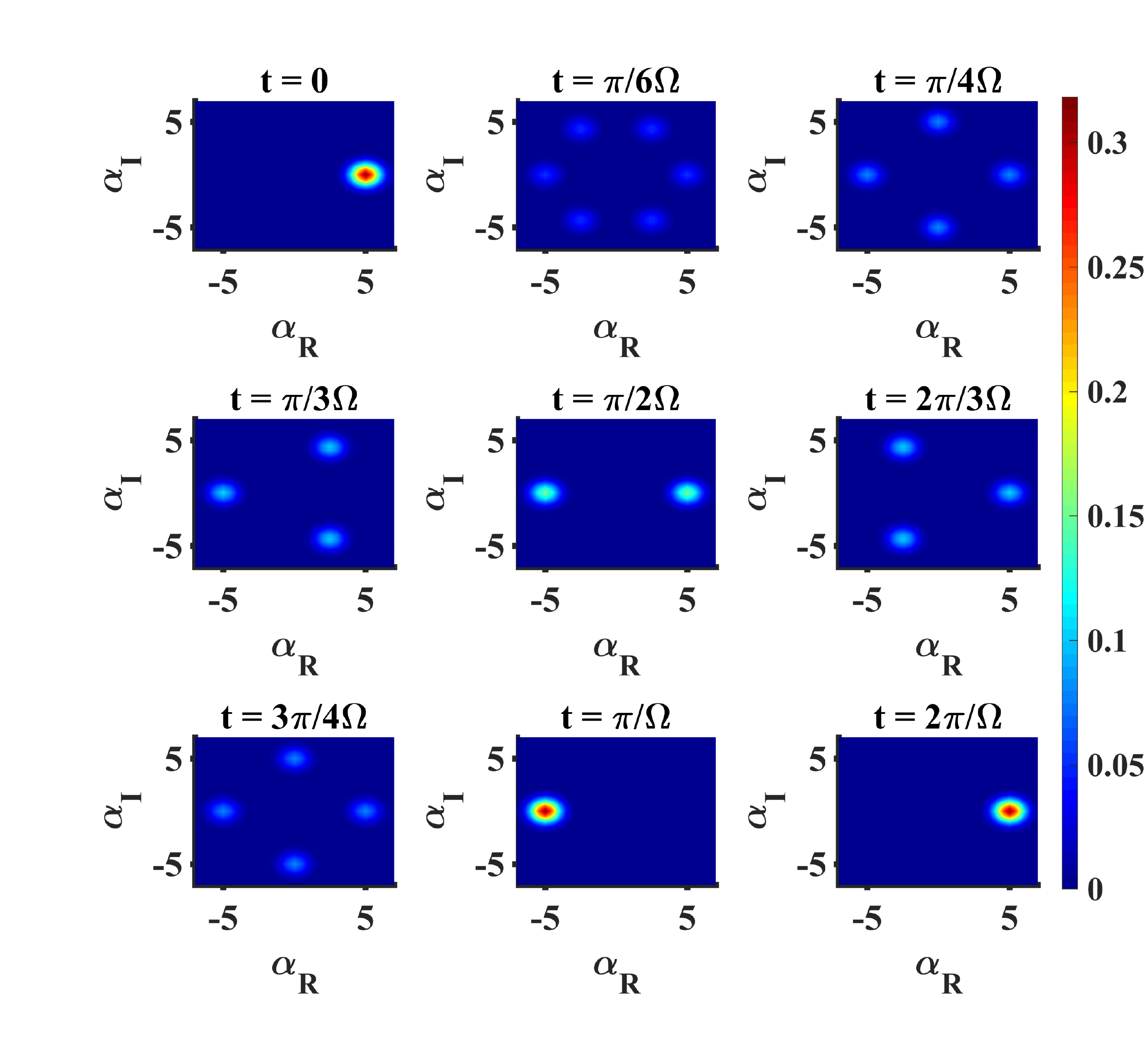}}

\textcolor{black}{}

\textcolor{black}{\caption{\label{fig:LGI(k=00003D2)}\textcolor{black}{The evolution of multi-component
cat-states for a system described by the nonlinear Hamiltonian eq.
(\ref{eq:ham}) where $k=2$. The contour graphs show su}rface plots
of $Q(\alpha)$ \textcolor{black}{with $\alpha_{0}=5$ for the times
$t=0$, $t=\pi/6\Omega$, $t=\pi/4\Omega$, $t=\pi/3\Omega$, $t=\pi/2\Omega$,
$t=2\pi/3\Omega$ and $t=3\pi/4\Omega$, $t=\pi/\Omega$ and $t=2\pi/\Omega$.}\textcolor{red}{{}
}}
}
\end{figure}

\textcolor{black}{}
\begin{figure*}[t]
\textcolor{black}{\includegraphics[width=2.1\columnwidth]{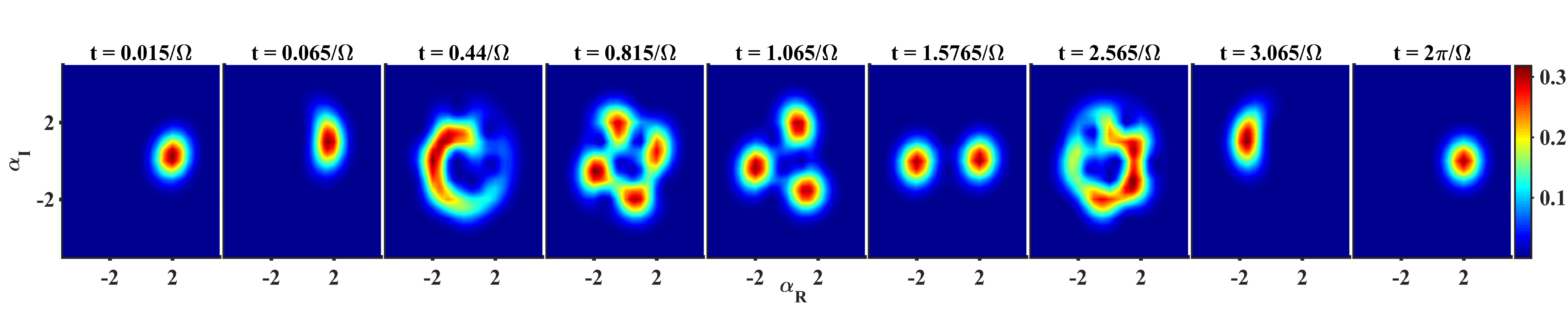}}

\textcolor{black}{\caption{The time sequence showing the evolution of multi-component cat-states
corresponding to the experiment of Kirchmair et al. \cite{collapse-revival-super-circuit}.
Here we take $\Omega$ to be negative, and $\alpha_{0}=2$ and $k=2$
for comparison with their reported results. We note the correspondence
$\Omega=-\Omega_{K}/2$ where $\Omega_{K}$ is the nonlinearity\textcolor{red}{{}
}defined by the Hamiltonian used in Ref. \cite{collapse-revival-super-circuit}.
The entire sequence was observed by them, except for the tri-cat state
at $t=2\pi/3$. Here we show the surface plots of the $Q$ functions
of the state at the given times.\textcolor{blue}{}}
}
\end{figure*}

\textcolor{black}{}
\begin{figure}[h]
\textcolor{black}{}

\textcolor{black}{\includegraphics[width=1\columnwidth]{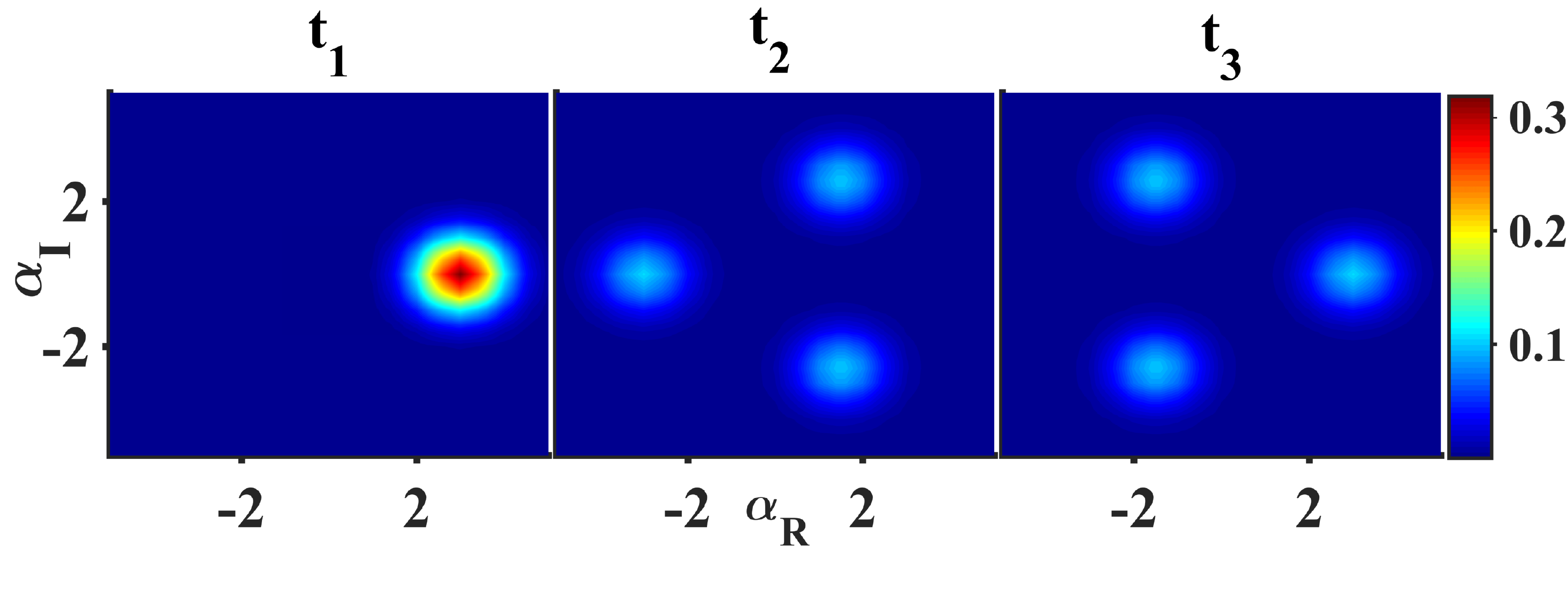}}

\textcolor{black}{\caption{Surface plot of $Q(\alpha)$ \textcolor{black}{with $\alpha_{0}=3$
and $k=2$ for the times $t=0$, $t=\pi/3\Omega$ and $t=2\pi/3\Omega$.
}}
}
\end{figure}
\textcolor{black}{Specifically, we will consider the times $t_{1}=0$,
$t_{2}=\pi/3\Omega$ and $t_{3}=2\pi/3\Omega$. This sequence for
$\alpha=3$ is plotted in Figure 6. To give violation of the LG inequality,
we select the values of $S_{i}$ differently at each of the times.
This does not affect the derivation of the inequality, as shown in
Section II. We define $S_{1}=+1$ if $x\geq0$ and $S_{1}=-1$ otherwise.
Similarly, we define $S_{2}=+1$ if $x\geq0$ and $S_{2}=0$ otherwise.
Finally, we define $S_{3}=-1$ if $x\leq0$ and $S_{3}=0$ otherwise.
Proceeding with the evaluation of the necessary moments, we find for
$\alpha\geq2$ on integrating $P(x)$ that $\begin{array}{ccc}
\langle S_{1}S_{2}\rangle & = & \frac{2}{3}\end{array}$ and $\begin{array}{ccc}
\langle S_{1}S_{3}\rangle & = & -\frac{2}{3}\end{array}$ (refer Table 1). }

\textcolor{black}{}
\begin{table}[H]
\textcolor{black}{}%
\begin{tabular}{|c|c|}
\hline 
\textcolor{black}{$t_{2}$} & \textcolor{black}{$t_{3}$}\tabularnewline
\hline 
\hline 
\textcolor{black}{$\begin{array}{ccc}
x\geq0, & S_{2}=+1, & P_{+}=\frac{2}{3}\\
x<0, & S_{2}=0, & P_{-}=\frac{1}{3}
\end{array}$} & \textcolor{black}{$\begin{array}{ccc}
x>0, & S_{3}=0, & P_{+}=\frac{1}{3}\\
x\leq0, & S_{3}=-1, & P_{-}=\frac{2}{3}
\end{array}$}\tabularnewline
\hline 
\end{tabular}\textcolor{black}{\caption{Table showing the probabilities for outcomes at times $t_{2}$ and
$t_{3}$ as evaluated for $\alpha\geq2$.\textcolor{red}{}}
}
\end{table}

\textcolor{black}{To evaluate $\langle S_{2}S_{3}\rangle$, we follow
the ``measure and re-prepare'' approach explained in Sections II
and IV. The expansion of the state at time $t_{2}$ is given as 
\begin{eqnarray}
|\psi\rangle & = & i\frac{1}{\sqrt{3}}\left|-\alpha\right\rangle +\frac{1}{\sqrt{3}}\exp(-i\pi/6)\left|e^{i\pi/3}\alpha\right\rangle \nonumber \\
 &  & +\frac{1}{\sqrt{3}}\exp(-i\pi/6)\left|e^{-i\pi/3}\alpha\right\rangle \label{eq:tricat}
\end{eqnarray}
This state is evident by the $Q$ function given in the plot of Figure
6. At time $t_{2}$, the system is thus in a superposition 
\begin{equation}
|\psi\rangle=N_{0}(|\psi_{1}\rangle+|\psi_{2}\rangle)\label{eq:sup}
\end{equation}
where $|\psi_{1}\rangle=|-\alpha\rangle$ and}

\textcolor{black}{
\begin{equation}
\left|\psi_{2}\right\rangle =N_{2}\{\left|e^{i\pi/3}\alpha\right\rangle +\left|e^{-i\pi/3}\alpha\right\rangle \}\label{eq:psi2}
\end{equation}
The normalization constants are $\begin{array}{ccc}
{\color{black}N_{0}^{-2}} & {\color{black}=} & {\color{black}3\{1+2\exp(-\frac{3}{2}\alpha^{2})\cos(\frac{\sqrt{3}}{2}\alpha^{2})\}}\end{array}$ and ${\color{black}N_{2}^{-2}=2\{1+\exp(-|\alpha|{}^{2}-\frac{1}{2}\alpha^{2})\cos(\frac{\sqrt{3}}{2}\alpha^{2})\}}$),
noting the initial condition implies $\alpha$ real. The assumption
of the macro-realism is that the system is }\textcolor{black}{\emph{in
one or the other}}\textcolor{black}{{} of the states $|\psi_{1}\rangle$
and $|\psi_{2}\rangle$ at time $t_{2}$. Using equation (\ref{eq:smix}),
we thus evaluate $\langle S_{2}S_{3}\rangle=P_{+}^{(2)}\langle S_{2}S_{3}\rangle_{+}+P_{-}^{(2)}\langle S_{2}S_{3}\rangle_{-}$
as given by Eq. (\ref{eq:smix}), where $P_{-}^{(2)}$ is the probability
the system is in state $|\psi_{1}\rangle$ at time $t_{2}$, and $\langle S_{2}S_{3}\rangle_{-}$
is the two-time moment given the system is in the state $|\psi{}_{1}\rangle$
at time $t_{2}$. Similarly, $P_{+}^{(2)}$ is the probability the
system is in state $|\psi_{2}\rangle$ at time $t_{2}$, and $\langle S_{2}S_{3}\rangle_{+}$
is the two-time moment given the system is in the state $|\psi_{2}\rangle$
at time $t_{2}$. Following the ``measure and re-prepare'' procedure
to evaluate $\langle S_{2}S_{3}\rangle_{-}$, we first assume the
system }\textcolor{black}{\emph{was}}\textcolor{black}{{} in $|\psi_{1}\rangle=|-\alpha\rangle$
at time $t_{2}=\pi/3\Omega$. We then evaluate what would have been
the state at the later time $2\pi/3\Omega$, after evolution for a
time $t=\pi/3\Omega$, with the initial state being $|\psi_{1}\rangle$.
The final state in this case is the tri-cat state depicted in Figures
4 and Figure 6 at $t=2\pi/3\Omega$. From our definitions of $S_{2}$
and $S_{3}$ (Table 1), we see that $\langle S_{2}S_{3}\rangle_{-}=0$.}

\textcolor{black}{}
\begin{figure}[h]
\textcolor{black}{\includegraphics[width=0.5\columnwidth]{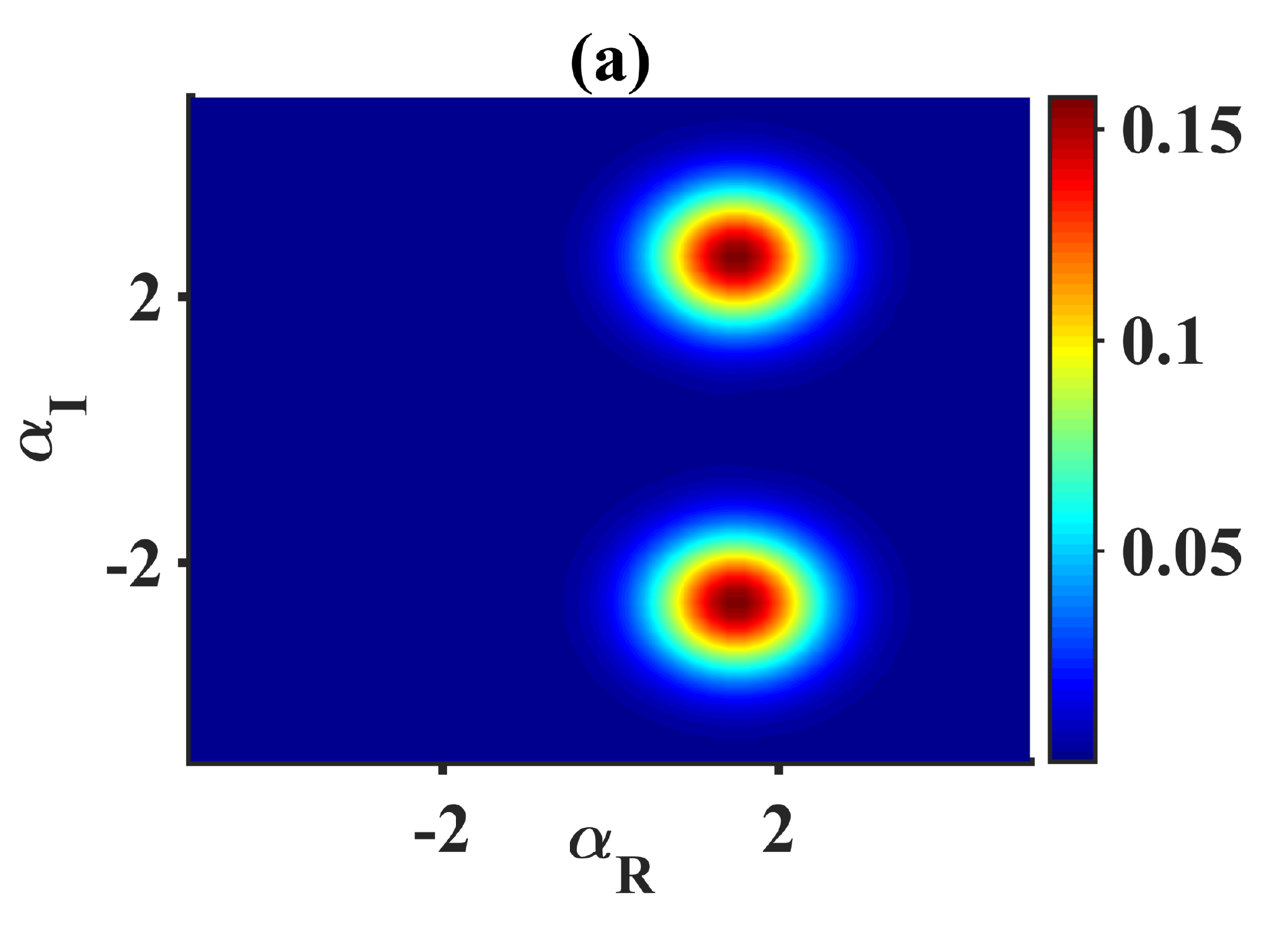}\includegraphics[width=0.5\columnwidth]{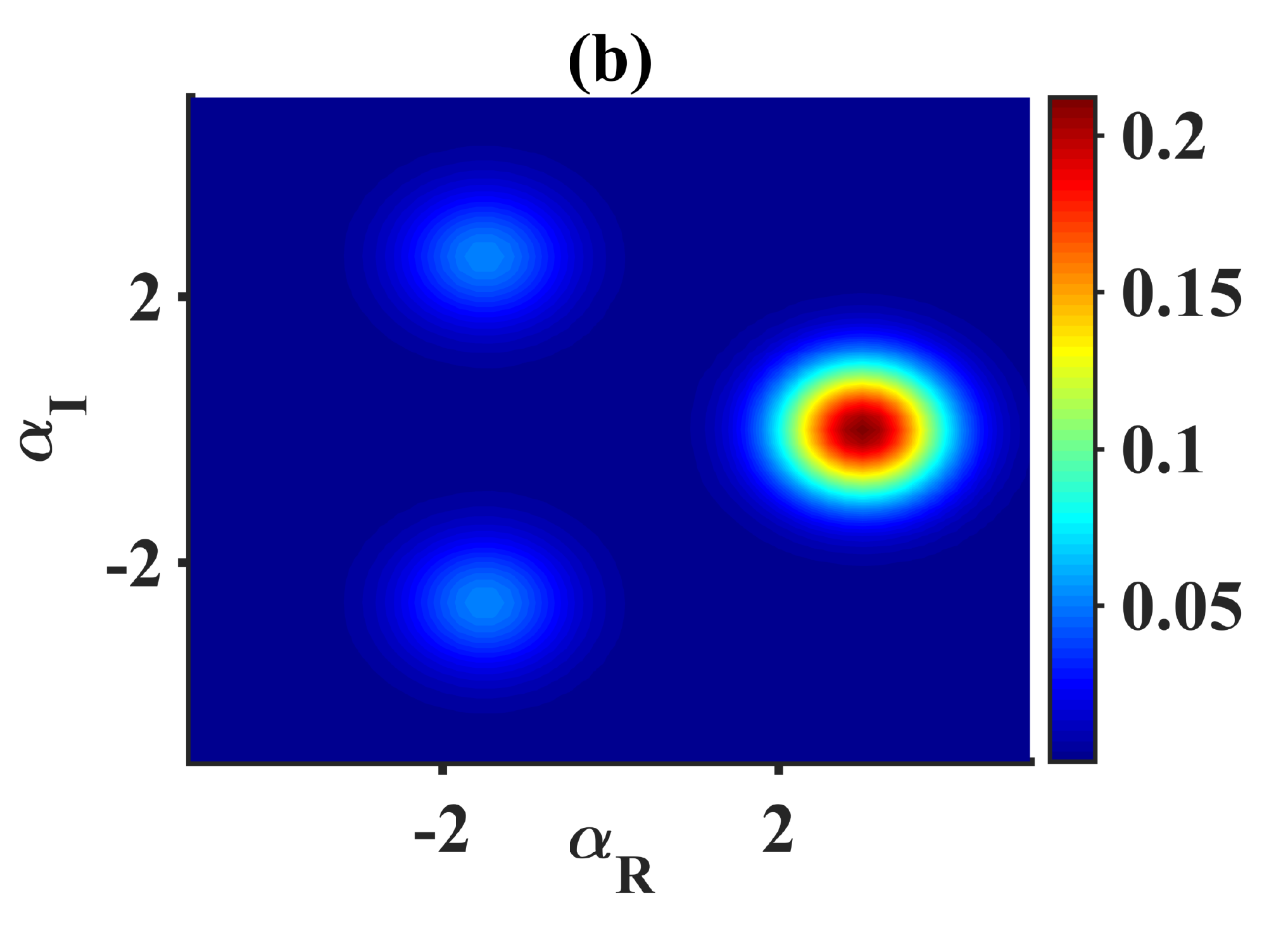}}

\textcolor{black}{\caption{Surface plot of $Q(\alpha)$ \textcolor{black}{with $\alpha_{0}=3$
and $k=2$. Figure (a) represents the initial state }given by equation
(\ref{eq:psi2}). Figure (b) is the time evolved state formed after
a time $\pi/3\Omega$, as\textcolor{black}{{} given by equation (\ref{eq:state}).}\textcolor{blue}{}\textcolor{red}{}}
}
\end{figure}

\textcolor{black}{}
\begin{figure}[t]
\textcolor{black}{\includegraphics[width=0.8\columnwidth]{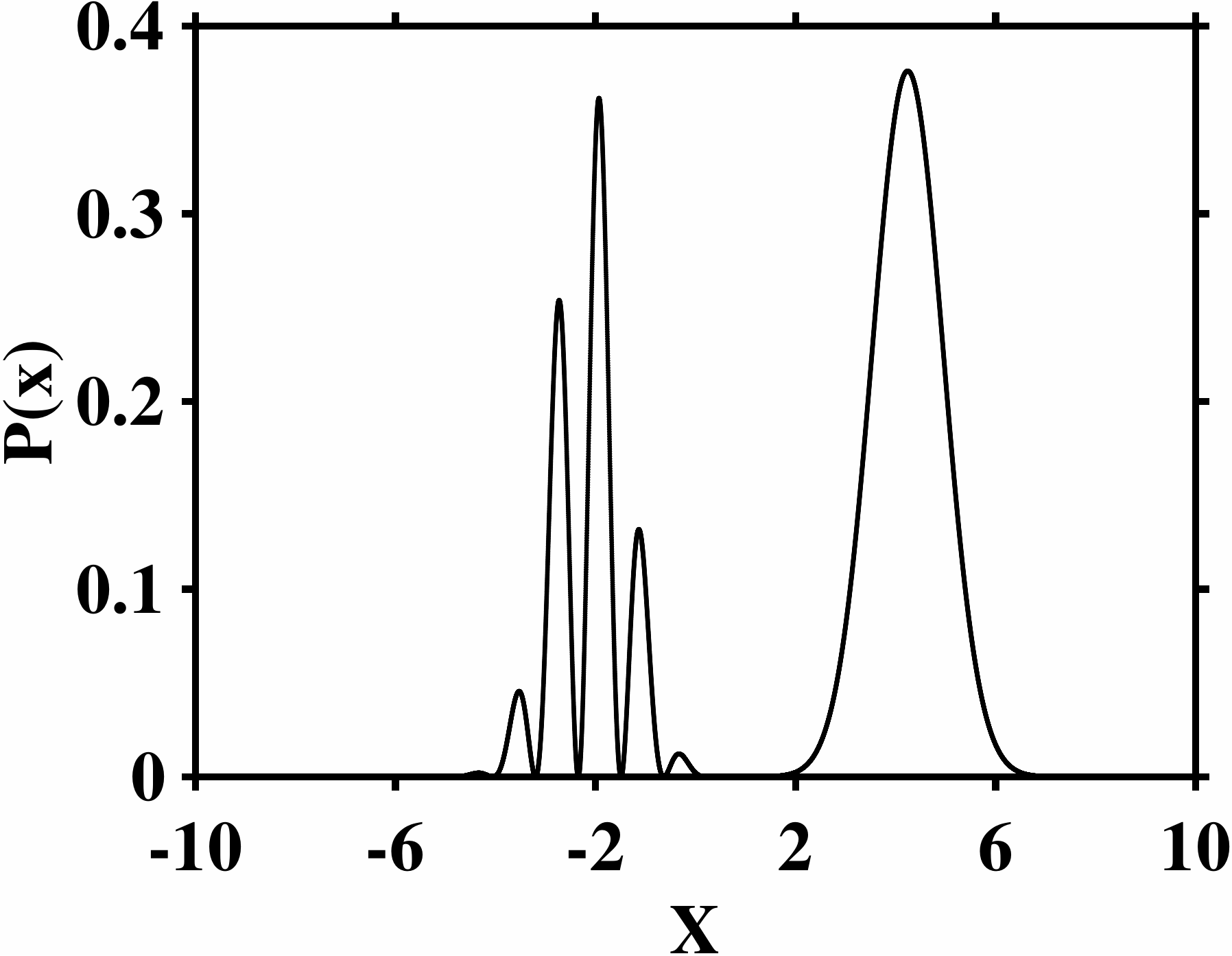}}

\textcolor{black}{\caption{\textcolor{black}{Plot of $P(x)$ for the state eq. (\ref{eq:state})
depicted in Figure 7b. The fringe pattern on the left is due to the
superposition of the coherent states $|-e^{i\pi/3}\alpha\rangle$
and $|-e^{-i\pi/3}\alpha\rangle$. The probability of obtaining positive
and negative results for $x$ are found on integration to be $P_{+}=0.6669$
and $P_{-}=0.3331$.}\textcolor{red}{{} }\textcolor{blue}{}}
}
\end{figure}
\textcolor{black}{Continuing with the evaluation of $\langle S_{2}S_{3}\rangle$
based on the ``measure and re-prepare'' strategy, we next assume
the system }\textcolor{black}{\emph{was}}\textcolor{black}{{} in $|\psi_{2}\rangle$
at the time $t_{2}$. This state is depicted by its $Q$ function
in Figure 7a. We then take this state to be the initial state for
a calculation where the system evolves for a time $\pi/3\Omega$,
to obtain the state that would have been generated at the time $t_{3}=2\pi/3\Omega$
with initial state $|\psi_{2}\rangle$. The state generated is}

\textcolor{black}{
\begin{eqnarray}
\left|\psi,\pi/3\right\rangle _{2} & = & N[\frac{2}{\sqrt{3}}e^{-i\pi/6}\left|\alpha\right\rangle \nonumber \\
 &  & +\frac{1}{3}(2-e^{-i\pi/3})\{\left|-e^{i\pi/3}\alpha\right\rangle +\left|-e^{-i\pi/3}\alpha\right\rangle \}]\nonumber \\
\label{eq:state}
\end{eqnarray}
where $N^{-2}=2[1+\exp(-\frac{3}{2}\alpha^{2})\cos(\frac{\sqrt{3}}{2}\alpha^{2})]$.
This state is depicted by its $Q$ function in Figure 7b. One evaluates
the $P(x)$ for (\ref{eq:state}), to find }

\textcolor{black}{
\begin{eqnarray}
P(x) & = & \frac{N^{2}}{\sqrt{\pi}}\exp(-x^{2})\Bigl\{\frac{4}{3}\exp(2\sqrt{2}x\alpha-2\alpha{}^{2})\nonumber \\
 &  & +\frac{4}{3}\exp(\frac{x\alpha}{\sqrt{2}}-\frac{5\alpha{}^{2}}{4})\cos(\frac{x\alpha\sqrt{6}}{2}+\frac{\alpha^{2}\sqrt{3}}{4})\nonumber \\
 &  & +\frac{2}{3}\exp(-\sqrt{2}x\alpha-\frac{|\alpha|^{2}}{2})[1\nonumber \\
 &  & \ \ \ \ \ \ \ \ \ +\cos(\sqrt{6}x\alpha+\frac{\alpha^{2}\sqrt{3}}{2})]\Bigl\}\label{eq:px}
\end{eqnarray}
as plotted in Figure 8. Evaluation of integrals gives probabilities
of $P_{+}=2/3$ and $P_{-}=1/3$ for obtaining a positive and negative
result for measurement of $x$ on this state (Figure 8). Thus 
\begin{eqnarray}
\langle S_{2}S_{3}\rangle_{+} & = & P_{+}\langle S_{3}\rangle_{+}+P_{-}\langle S_{3}\rangle_{-}=-\frac{1}{3}\label{eq:sa2}
\end{eqnarray}
Using eq. (\ref{eq:smix}), we find  $\langle S_{2}S_{3}\rangle=-\frac{2}{9}$.
This implies}

\textcolor{black}{
\begin{equation}
\langle S_{1}S_{2}\rangle-\langle S_{1}S_{3}\rangle+\langle S_{2}S_{3}\rangle=10/9\label{eq:lg2}
\end{equation}
A violation of the Leggett-Garg inequality is obtained for the $k=2$
case.}

\section{Experimental strategy}

\textcolor{black}{Various experimental strategies can be used to demonstrate
violation of the Leggett-Garg inequalities. These are documented in
the literature \cite{lgexpphotonweak,legggarg,laura-lg-two-well,jordan_kickedqndlg2,experiment-lg,Mitchell,stat-1zhou,stat-2-neutrino,small-systemlg,emeryreview,massiveosci,NSTmunro}.
The inequality involves two-time correlation functions of three observables,
$S_{1}$, $S_{2}$ and $S_{3}$, measured at three different times.
A common strategy is to evaluate the two-time correlation moments
by taking an ensemble average of an appropriate two-time moment with
suitable initial states. As applied to this proposal, for two of the
moments ($\langle S_{1}S_{2}\rangle$ and $\langle S_{1}S_{3}\rangle$)
the initial state at the time $t_{1}$ is a coherent state, identical
to the experiments \cite{collapse-revival-bec,collapse-revival-super-circuit}.
The state formed at the intermediate time $t_{2}$ is a superposition
of two states $\psi_{1}$ and $\psi_{2}$, which are well-separated
and distinguishable by a measurement of a quadrature amplitude $\hat{x}$,
as defined in Sections IV and V. The sign of the outcome for $\hat{x}$
determines the value of $S_{2}$. The moments obtained experimentally
if this measurement $\hat{x}$ were performed can be evaluated from
the experimentally determined $Q$ functions, given in Refs. \cite{collapse-revival-super-circuit}.
The validity of the evaluation of the moments for a test of macro-realism
is based on the validity of the macro-realism premises. It is assumed
that the system prior to any measurement will be in one of several
macroscopically distinct states available to it. The value $S_{1}$
at time $t_{1}$ is known by preparation: the coherent state at time
$t_{1}$ is prepared as $|\alpha\rangle$, which has a positive value
for outcome $\hat{x}$ so that $S_{1}=1$. The value $S_{3}$ or $S_{2}$
(for evaluation of $\langle S_{1}S_{2}\rangle$ and $\langle S_{1}S_{3}\rangle$)
is measurable by a projective measurement of $\hat{x}$ and hence
of $S_{3}$ (or $S_{2}$): It is assumed this measurement correctly
measures which of the states the system was in (prior to the measurement)
to a macroscopic level of precision $-$ which is all that is necessary
to determine $S_{i}$ prior to the measurement. In evaluating $\langle S_{1}S_{3}\rangle$,
the measurement at time $t_{2}$ does not need to be made.}

\textcolor{black}{The moment $\langle S_{2}S_{3}\rangle$ is measurable
using different strategies, including weak and ideal negative-result
measurements, as discussed in Refs. \cite{lgexpphotonweak,legggarg,laura-lg-two-well,jordan_kickedqndlg2,experiment-lg,Mitchell,stat-1zhou,stat-2-neutrino,small-systemlg,emeryreview,massiveosci,NSTmunro}.
Here, we suggest a simple approach, as in Refs. \cite{NSTmunro,stat-1zhou,stat-2-neutrino,laura-lg-two-well}.
It can be verified experimentally that at time $t_{2}$, the system
is in the superposition of two macroscopically distinguishable states
$\psi_{1}$ and $\psi_{2}$. The moment $\langle S_{2}S_{3}\rangle$
can be measured by first preparing the system in one state $\psi_{1}$
and then the other $\psi_{2}$, and measuring the moment $\langle S_{2}S_{3}\rangle$
for each case. The final moment is evaluated from the weighted average,
assuming a }\textcolor{black}{\emph{stationarity}}\textcolor{black}{,
that the system evolves similarly under time translations \cite{stat-1zhou,stat-2-neutrino}.
If the system is indeed in one or other state at time $t_{2}$ (as
the first macro-realism premise implies), then the measured moment
is justified to be the value $\langle S_{2}S_{3}\rangle$ that would
be measured, if the ideal macroscopically noninvasive measurement
could take place. A violation of the Leggett-Garg inequality observed
with this approach then serves to invalidate the premise of macro-realism. }

\textcolor{black}{To carry out the evaluation of $\langle S_{2}S_{3}\rangle$
for $k=2$, we note that the first state $\psi_{1}$ of the superposition
formed at time $t_{2}$ is a coherent state, which can be prepared
and evolved to the tri-cat as in Ref \cite{collapse-revival-super-circuit}.
The second state $\psi_{2}$ of the superposition is itself a superposition,
of two coherent states $\pi$ out of phase. This state $\psi_{2}$
can be re-prepared, up to a rotation in phase space, by evolving the
coherent state for a time $t=\pi/2\Omega$, as illustrated in Figure
4. Evidence for the generation of this state $\psi_{2}$ is given
in the BEC and superconducting circuit experiments of Refs. \cite{collapse-revival-bec,collapse-revival-super-circuit}.
That this state is indeed generated can be established via tomography
using the $Q$ function. In the anticipated experiment, the re-prepared
state $\psi_{2}$ is then evolved for a time corresponding to $\pi/3\Omega$.
The predicted state after this evolution is given in Figure 6b, from
which the moments $\langle S_{2}S_{3}\rangle$ can be evaluated as
described in Section V. The re-prepared state used here is different
to $\psi_{2}$ by a rotation in phase space. One can justify that
the measured correlation $\langle S_{2}S_{3}\rangle$ is unchanged,
assuming the invariance of moments under rotations in phase space.
Alternatively, one can experimentally obtain the rotated state, by
rotating the initial coherent state.}

\section{Conclusion}

\textcolor{black}{In summary, we have derived generalizations of Leggett-Garg
inequalities and demonstrated how the new inequalities can be used
to test macro-realism for dynamical cat-states created by a nonlinearity.
In particular, in Section V we demonstrate the feasibility of violating
a Leggett-Garg inequality using a Kerr $\chi^{(3)}$ nonlinearity.
This enables us to predict violation of Leggett-Garg inequalities
for the experiments of Greiner et al. \cite{collapse-revival-bec}
and Kirchmair et al. \cite{collapse-revival-super-circuit}. These
experiments observe the collapse and revival of a coherent state over
the necessary timescales. }

\textcolor{black}{Finally, we comment on loopholes and on the significance
of the proposed Leggett-Garg test. First, we need to assume that the
system at time $t_{1}$ is reliably prepared in the coherent state,
and that the measurement at time $t_{3}$ accurately records the value
of $S_{3}$ of the state of the system prior to measurement. The violation
of the inequality would then negate that the system is in one or other
of the macroscopically distinguishable states $\psi_{1}$ and $\psi_{2}$
at the time $t_{2}$ (or similarly at time $t_{3}$). This is because
a system in a classical mixture $\rho_{mix}$ of these states at these
times would satisfy the Leggett-Garg premises. (This is understood
because the system is in one of the two macroscopically distinguishable
states, and a measurement can be constructed that leaves these states
unchanged). Hence if the system is in a classical mixture $\rho_{mix}$
of the two states, one could }\textcolor{black}{\emph{not}}\textcolor{black}{{}
generate a violation of the inequalities. The violation of the Leggett-Garg
inequality thus gives a demonstration of a mesoscopic quantum coherence.}

\textcolor{black}{It could be argued that the failure of the mixture
$\rho_{mix}$ is also exemplified by the observation of the revival
of the final coherent state as observed in the experiments of Refs.
\cite{collapse-revival-super-circuit,collapse-revival-bec}, since
a classical mixture $\rho_{mix}$ would not give a such a revival.
However, this latter argument does not rule out that the system might
be describable as another mixture consistent with the macro-realism
premises in a theory alternative to quantum mechanics. In this respect,
the violation of the Leggett-Garg inequality, which does not rely
on quantum mechanics, is designed to give a stronger conclusion.}

\textcolor{black}{Extra assumptions also exist for the Leggett-Garg
test however, which create loopholes. The proposed strategy relies
on the preparation of the states $\psi_{1}$ and $\psi_{2}$ at the
time $t_{2}$ for the evaluation of the $\langle S_{2}S_{3}\rangle$,
and we thus assume the state regenerated for the later measurement
is the actual $\psi_{i}$. The objective of the Leggett-Garg inequality
however is to falsify the premises of macro-realism for all}\textcolor{black}{\emph{
}}\textcolor{black}{theories, not only quantum mechanics. For this
there is a potential loophole, since it could not be excluded that
for an alternative theory, the system at time $t_{2}$ is in a state
}\textcolor{black}{\emph{microscopically}}\textcolor{black}{{} different
to the state $\psi_{1}$ or $\psi_{2}$. These states may be minimally
disturbed by the measurement performed at $t_{2}$, and it might be
argued that this minimal disturbance may generate after evolution
a macroscopic change to the state (and hence to $S_{3}$) at the later
time $t_{3}.$ This would not affect the justification of the evaluation
of $\langle S_{2}S_{3}\rangle$ or $\langle S_{1}S_{3}\rangle$. However,
for the evaluation of $\langle S_{2}S_{3}\rangle$ one could then
not exclude that a small difference to the state measured at $t_{2}$
results in a macroscopic difference to the outcome at the time $t_{3}$.
To eliminate such a loophole, one is left with the difficult task
to regenerate all states that are microscopically different to $\psi_{1}$
and $\psi_{2}$, and demonstrate that the evolution from time $t_{2}$
to $t_{3}$ does not change the value of $S$ measured at $t_{3}$.
Alternatively, one can seek to perform the Leggett-Garg test using
an ideal negative-result measurement or a weak measurement \cite{legggarg,emeryreview}.}
\begin{acknowledgments}
\textcolor{black}{This research has been supported by the Australian
Research Council Discovery Project Grants schemes under Grant DP180102470.
This work was performed in part at Aspen Center for Physics, which
is supported by National Science Foundation grant PHY-1607611.}
\end{acknowledgments}

\section*{Appendix}

\subsubsection*{Coherent state expansion of states}

In the problems treated, we consider the initial state of the system
to be a coherent state $\left|\psi,t=0\right\rangle =\left|\alpha\right\rangle $
where $\alpha$ is real. In some cases, it is well-known that the
state created at a later time can be written as a superposition of
a finite number of distinct coherent states. A summary of some such
examples in given in the Table \ref{tab:psi}. Important for this
paper is that the state $|\psi,t=\pi/3\Omega\rangle$ (for $k=2$)
is a superposition of three coherent states $\left|-\alpha\right\rangle $,
$\left|e^{i\pi/3}\alpha\right\rangle $ and $\left|e^{-i\pi/3}\alpha\right\rangle $.
That this is so is suggested by the plot of the $Q$ function, given
in Figure \ref{fig:LGI(k=00003D2)}. A simple analysis allows us to
evaluate the probability amplitudes, given as $|\psi,t=\pi/3\Omega\rangle=A|-\alpha\rangle+B|e^{i\pi/3}\alpha\rangle+C|e^{-i\pi/3}\alpha\rangle$
. One can write the state at this time as

\begin{eqnarray*}
\left|\alpha,\pi/3\varOmega\right\rangle  & = & e^{-|\alpha|^{2}/2}\underset{n}{\sum}A(-1)^{n}\alpha^{n}\frac{1}{\sqrt{n!}}\left|n\right\rangle \\
 &  & +e^{-|\alpha|^{2}/2}\underset{n}{\sum}B(e^{i\pi/3})^{n}\alpha^{n}\frac{1}{\sqrt{n!}}\left|n\right\rangle \\
 &  & +e^{-|\alpha|^{2}/2}\underset{n}{\sum}C(e^{-i\pi/3})^{n}\alpha^{n}\frac{1}{\sqrt{n!}}\left|n\right\rangle \\
 & = & e^{-|\alpha|^{2}/2}\underset{n}{\sum}\exp(-i\pi n^{2}/3)\alpha^{n}\frac{1}{\sqrt{n!}}\left|n\right\rangle 
\end{eqnarray*}
Evaluation of the amplitudes $A$, $B$ and $C$ is done by simultaneously
solving for the coefficients, where $n=0,1,2,...$. By solving we
can write the wavefunction as in eq. (\ref{eq:tricat}). \textcolor{red}{}

\subsubsection*{\textcolor{black}{Evaluation of $P(x)$ and the $Q$ function}}

\textcolor{red}{}We evaluate $P(x)$ using the expression $P(x_{\theta})$
for the probability measurement: $P(x_{\theta})=|\langle x_{\theta}|\psi\rangle|^{2}$
where $\hat{x_{\theta}}=(e^{i\theta}\hat{\alpha}+e^{-i\theta}\hat{a^{\dagger}})\sqrt{2}$.
The generalized position representation $\psi_{\alpha,t}(x)=\left\langle x|\alpha,t\right\rangle $
can be written as \cite{yurke-stoler}
\[
\psi_{\alpha}(x_{\theta})=\frac{1}{\pi^{1/4}}\exp(-\frac{x^{2}}{2}+\frac{2x|\alpha|e^{i(\theta+\phi)}}{\sqrt{2}}-\frac{|\alpha|^{2}e^{2i(\theta+\phi)}}{2}-\frac{|\alpha|^{2}}{2})
\]
where $\alpha=|\alpha|e^{i\phi}$. By using the expression for $\theta=0$,
we see that $P(x)=\frac{1}{\pi^{1/2}}\exp(-x^{2}+2\sqrt{2}x\alpha-2|\alpha|^{2})$.
A summary of the position probability distributions $P(x)$ evaluated
for the various times is given in the Table \ref{tab:Px}. 

We also evaluate the Husimi Q \cite{Husimi-Q} representation defined
as\textcolor{black}{{} $Q(\alpha)=\langle\alpha|\rho|\alpha\rangle/\pi=\langle\alpha|\psi\rangle\langle\psi|\alpha\rangle/\pi=|\langle\alpha|\psi\rangle|^{2}/\pi$}.
At $t=0$, we take $\psi=|\alpha_{0}\rangle$. Then 
\[
Q(\alpha)=\frac{1}{\pi}\exp(-|\alpha|^{2}-|\alpha_{0}|^{2}+\alpha_{0}(\alpha^{*}+\alpha))
\]
 where we have considered $\alpha_{0}$ is real. The remaining functions
are calculated using that the inner product of two coherent states
is $\langle\alpha|\beta\rangle=\exp(-\frac{|\alpha|^{2}}{2}-\frac{|\beta|^{2}}{2}+\alpha^{*}\beta)$.
A summary of the functions $Q(\alpha)$ evaluated for various times
is given in the table \ref{tab:Q_alpha} below.

\onecolumngrid

\textcolor{black}{}
\begin{table}[H]
\noindent \begin{centering}
\textcolor{black}{}%
\begin{tabular}{|c|c|c|}
\hline 
 & \textcolor{black}{$\psi_{\alpha}$} & relevant $k$\tabularnewline
\hline 
\hline 
\textcolor{black}{$t=0$} & \textcolor{black}{$\left|\alpha\right\rangle $} & \textcolor{black}{all $k$}\tabularnewline
\hline 
\textcolor{black}{$t=\pi/8\Omega$ } & \textcolor{black}{$\begin{array}{c}
{\normalcolor \frac{\sqrt{2}}{4}e^{-i\pi/8}\{|e^{i\pi/4}\alpha\rangle-|-e^{i\pi/4}\alpha\rangle+}|e^{-i\pi/4}\alpha\rangle-|-e^{-i\pi/4}\alpha\rangle\}\\
+\frac{1}{4}\{(1-i)(|\alpha\rangle+|-\alpha\rangle)+(1+i)(|i\alpha\rangle+|-i\alpha\rangle)\}
\end{array}$} & \textcolor{black}{$k=2$}\tabularnewline
\hline 
\multirow{2}{*}{\textcolor{black}{$t=\pi/4\Omega$}} & \textcolor{black}{$\begin{array}{c}
\frac{1}{2}\{|-i\alpha\rangle+|i\alpha\rangle+e^{-i\pi/4}|\alpha\rangle-e^{-i\pi/4}|-\alpha\rangle\}\\
\\
\end{array}$} & \textcolor{black}{$\begin{array}{c}
k=2\\
\\
\end{array}$}\tabularnewline
\cline{2-3} 
 & $\frac{1}{2}\{(1+e^{-i\pi/4})|\alpha\rangle+(1-e^{-i\pi/4})|-\alpha\rangle\}$ & $k>2$, $k$ even\tabularnewline
\hline 
\textcolor{black}{$t=\pi/3\Omega$} & $i\frac{1}{\sqrt{3}}\left|-\alpha\right\rangle +\frac{1}{\sqrt{3}}\exp(-i\pi/6)\left|e^{i\pi/3}\alpha\right\rangle +\frac{1}{\sqrt{3}}\exp(-i\pi/6)\left|e^{-i\pi/3}\alpha\right\rangle $ & $k=2$\tabularnewline
\hline 
\textcolor{black}{$t=3\pi/8\Omega$ } & \textcolor{black}{$\begin{array}{c}
{\normalcolor \frac{\sqrt{2}}{4}e^{-i3\pi/8}\{|e^{i\pi/4}\alpha\rangle-|-e^{i\pi/4}\alpha\rangle+}|e^{-i\pi/4}\alpha\rangle-|-e^{-i\pi/4}\alpha\rangle\}\\
+\frac{1}{4}\{(1+i)(|\alpha\rangle+|-\alpha\rangle)+(1-i)(|i\alpha\rangle+|-i\alpha\rangle)\}
\end{array}$} & \textcolor{black}{$k=2$}\tabularnewline
\hline 
\multirow{2}{*}{\textcolor{black}{$t=\pi/2\Omega$}} & \textcolor{black}{$\begin{array}{c}
\frac{1}{\sqrt{2}}(e^{-i\frac{\pi}{4}}|\alpha\rangle+e^{+i\frac{\pi}{4}}|-\alpha\rangle)\\
\\
\end{array}$} & \textcolor{black}{$k$ even}\tabularnewline
\cline{2-3} 
 & $\frac{1}{2}(|\alpha\rangle+|-\alpha\rangle+|-i\alpha\rangle-|i\alpha\rangle)$ & $k$ odd\tabularnewline
\hline 
\textcolor{black}{$t=5\pi/8\Omega$ } & \textcolor{black}{$\begin{array}{c}
{\normalcolor \frac{\sqrt{2}}{4}e^{-i5\pi/8}\{|e^{i\pi/4}\alpha\rangle-|-e^{i\pi/4}\alpha\rangle+}|e^{-i\pi/4}\alpha\rangle-|-e^{-i\pi/4}\alpha\rangle\}\\
+\frac{1}{4}\{(1-i)(|\alpha\rangle+|-\alpha\rangle)+(1+i)(|i\alpha\rangle+|-i\alpha\rangle)\}
\end{array}$} & \textcolor{black}{$k=2$}\tabularnewline
\hline 
\textcolor{black}{$t=2\pi/3\Omega$} & \textcolor{black}{$\frac{1-2e^{i\pi/3}}{3}\left|\alpha\right\rangle +\frac{1}{\sqrt{3}}e^{i\pi/6}\left|-e^{i\pi/3}\alpha\right\rangle +\frac{1}{\sqrt{3}}e^{i\pi/6}\left|-e^{-i\pi/3}\alpha\right\rangle )$} & \textcolor{black}{$k=2$}\tabularnewline
\hline 
\multirow{2}{*}{\textcolor{black}{$t=3\pi/4\Omega$}} & \textcolor{black}{$\begin{array}{c}
\frac{1}{2}\{|-i\alpha\rangle+|i\alpha\rangle-e^{+i\pi/4}|\alpha\rangle+e^{+i\pi/4}|-\alpha\rangle\}\\
\\
\end{array}$} & \textcolor{black}{$\begin{array}{c}
k=2\\
\\
\end{array}$}\tabularnewline
\cline{2-3} 
 & $\frac{1}{2}\{(1-e^{+i\pi/4})|\alpha\rangle+(1+e^{+i\pi/4})|-\alpha\rangle\}$ & $k>2$, $k$ even\tabularnewline
\hline 
\textcolor{black}{$t=7\pi/8\Omega$ } & \textcolor{black}{$\begin{array}{c}
{\normalcolor \frac{\sqrt{2}}{4}e^{-i7\pi/8}\{|e^{i\pi/4}\alpha\rangle-|-e^{i\pi/4}\alpha\rangle+}|e^{-i\pi/4}\alpha\rangle-|-e^{-i\pi/4}\alpha\rangle\}\\
+\frac{1}{4}\{(1+i)(|\alpha\rangle+|-\alpha\rangle)+(1-i)(|i\alpha\rangle+|-i\alpha\rangle)\}
\end{array}$} & \textcolor{black}{$k=2$}\tabularnewline
\hline 
\textcolor{black}{$t=\pi/\Omega$} & \textcolor{black}{$\left|-\alpha\right\rangle $} & \textcolor{black}{all $k$}\tabularnewline
\hline 
\end{tabular}
\par\end{centering}
\noindent \centering{}\textcolor{black}{\caption{\label{tab:psi}Summary of the multi-component cat-states formed at
different times, for different $k$.}
}
\end{table}

\begin{table}[H]
\begin{centering}
\textcolor{black}{}%
\begin{tabular}{|c|c|c|}
\hline 
 & \textcolor{black}{$P(x)$} & \textcolor{black}{$k$}\tabularnewline
\hline 
\hline 
\textcolor{black}{$t=0$} & \textcolor{black}{$\begin{array}{c}
\frac{1}{\pi^{1/2}}\exp(-x^{2}+2\sqrt{2}x\alpha-2|\alpha|^{2})\end{array}$} & \textcolor{black}{all $k$}\tabularnewline
\hline 
\multirow{2}{*}{\textcolor{black}{$t=\pi/4\Omega$}} & $\begin{array}{c}
\frac{1}{2}\frac{1}{\pi^{1/2}}\exp(-x^{2}-|\alpha|^{2})\{2\sinh(|\alpha|^{2})+\exp(-|\alpha|^{2})\cosh(2\sqrt{2}x\alpha)\\
+\exp(|\alpha|^{2})\cos(2\sqrt{2}x\alpha)+2\sqrt{2}\cos(\sqrt{2}x\alpha)\sinh(\sqrt{2}x\alpha)\}
\end{array}$ & \textcolor{black}{$k=2$}\tabularnewline
\cline{2-3} 
 & $\begin{array}{c}
\begin{array}{c}
\frac{1}{\pi^{1/2}}\exp(-x^{2}-2|\alpha|^{2})\{\cosh(2\sqrt{2}x\alpha)+\frac{1}{\sqrt{2}}\sinh(2\sqrt{2}x\alpha)\}\end{array}\end{array}$ & $k>2$, $k$ even\tabularnewline
\hline 
\textcolor{black}{$t=\pi/3\Omega$} & $\begin{array}{c}
\frac{1}{3}\frac{1}{\pi^{1/2}}\exp(-x^{2})\{\exp(-2\sqrt{2}x\alpha-2\alpha^{2})+2\exp(\sqrt{2}x\alpha-\alpha{}^{2}/2)\\
-2\exp(-\frac{\sqrt{2}}{2}x\alpha-5\alpha^{2}/4)\cos(\frac{\sqrt{6}}{2}x\alpha-\frac{\sqrt{3}}{4}\alpha^{2})\\
+2\exp(\sqrt{2}x\alpha-\alpha^{2}/2)\cos(\sqrt{6}x\alpha-\frac{\sqrt{3}}{2}\alpha^{2})\}
\end{array}$ & $k=2$\tabularnewline
\hline 
\textcolor{black}{$t=\pi/2\Omega$} & \textcolor{black}{$\frac{1}{\pi^{1/2}}\exp(-x^{2}-2|\alpha|^{2})\cosh(2\sqrt{2}xa)$ } & \textcolor{black}{$k$ even}\tabularnewline
\hline 
\multirow{2}{*}{\textcolor{black}{$t=3\pi/4\Omega$}} & $\begin{array}{c}
\frac{1}{2}\frac{1}{\pi^{1/2}}\exp(-x^{2}-|\alpha|^{2})\{2\sinh(|\alpha|^{2})+\exp(-|\alpha|^{2})\cosh(2\sqrt{2}x\alpha)\\
+\exp(|\alpha|^{2})\cos(2\sqrt{2}x\alpha)-2\sqrt{2}\cos(\sqrt{2}x\alpha)\sinh(\sqrt{2}x\alpha)\}
\end{array}$ & $k=2$\tabularnewline
\cline{2-3} 
 & $\begin{array}{c}
\frac{1}{\pi^{1/2}}\exp(-x^{2}-2|\alpha|^{2})\{\cosh(2\sqrt{2}x\alpha)\\
-\frac{1}{\sqrt{2}}\sinh(2\sqrt{2}x\alpha)\}
\end{array}$ & $k>2$, $k$ even\tabularnewline
\hline 
\textcolor{black}{$t=\pi/\Omega$} & \textcolor{black}{$\begin{array}{c}
\frac{1}{\pi^{1/2}}\exp(-x^{2}-2\sqrt{2}x\alpha-2|\alpha|^{2})\end{array})$} & \textcolor{black}{all $k$ }\tabularnewline
\hline 
\end{tabular}
\par\end{centering}
\caption{\label{tab:Px}Summary of the probability distributions $P(x)$ for
various $t$ and $k$.\textcolor{red}{}}
\end{table}

\begin{table}[H]
\begin{centering}
\textcolor{black}{}%
\begin{tabular}{|c|c|c|}
\hline 
\textcolor{black}{$t$} & $Q(\alpha)$ & \textcolor{black}{$k$}\tabularnewline
\hline 
\hline 
\textcolor{black}{$t=0$} & \textcolor{black}{$\frac{1}{\pi}\exp(-|\alpha|^{2}-|\alpha_{0}|^{2}+\alpha_{0}(\alpha^{*}+\alpha))$} & \textcolor{black}{all $k$}\tabularnewline
\hline 
\multirow{2}{*}{\textcolor{black}{$t=\pi/4\Omega$}} & $\begin{array}{c}
\frac{1}{\pi}\exp(-|\alpha|^{2}-|\alpha_{0}|^{2})\{\cos(\alpha^{*}\alpha_{0})+e^{-i\pi/4}\sinh(\alpha^{*}\alpha_{0})\}\\
\times\{\cos(\alpha_{0}\alpha)+e^{+i\pi/4}\sinh(\alpha_{0}\alpha)\}
\end{array}$ & \textcolor{black}{$k=2$}\tabularnewline
\cline{2-3} 
 & $\begin{array}{c}
\frac{1}{\pi}\exp(-|\alpha|^{2}-|\alpha_{0}|^{2})\{\cosh(\alpha\alpha_{0}+\alpha^{*}\alpha_{0})\\
+\frac{1}{\sqrt{2}}\sinh(\alpha\alpha_{0}+\alpha^{*}\alpha_{0})+i\frac{1}{\sqrt{2}}\sinh(\alpha\alpha_{0}-\alpha^{*}\alpha_{0})\}
\end{array}$ & $k>2$, $k$ even\tabularnewline
\hline 
\textcolor{black}{$t=\pi/3\Omega$} & $\begin{array}{c}
\frac{1}{3}\frac{1}{\pi}\exp(-|\alpha_{0}|{}^{2}-|\alpha|^{2}+\frac{1}{2}\alpha_{0}\alpha^{*}+\frac{1}{2}\alpha_{0}\alpha)\{\exp(-\frac{3}{2}\alpha\alpha_{0})+2ie^{i\pi/6}\cos(\frac{\sqrt{3}}{2}\alpha\alpha_{0})\}\\
\times\{\exp(-\frac{3}{2}\alpha^{*}\alpha_{0})-2ie^{-i\pi/6}\cos(\frac{\sqrt{3}}{2}\alpha^{*}\alpha_{0})\}
\end{array}$ & $k=2$\tabularnewline
\hline 
\textcolor{black}{$t=\pi/2\Omega$} & \textcolor{black}{$\begin{array}{c}
\frac{1}{\pi}\exp(-|\alpha|^{2}-|\alpha_{0}|^{2})[\cosh(\alpha^{*}\alpha_{0}+\alpha\alpha_{0})-i\sinh(\alpha^{*}\alpha_{0}-\alpha\alpha_{0})]\end{array}$} & \textcolor{black}{$k$ even}\tabularnewline
\hline 
\multirow{2}{*}{\textcolor{black}{$t=3\pi/4\Omega$}} & $\begin{array}{c}
\frac{1}{\pi}\exp(-|\alpha|^{2}-|\alpha_{0}|^{2})\{\cos(\alpha^{*}\alpha_{0})-e^{+i\pi/4}\sinh(\alpha^{*}\alpha_{0})\}\\
\times\{\cos(\alpha_{0}\alpha)-e^{-i\pi/4}\sinh(\alpha_{0}\alpha)\}
\end{array}$ & $k=2$\tabularnewline
\cline{2-3} 
 & $\begin{array}{c}
\frac{1}{\pi}\exp(-|\alpha|^{2}-|\alpha_{0}|^{2})\{\cosh(\alpha^{*}\alpha_{0}+\alpha\alpha_{0})\\
-\frac{1}{\sqrt{2}}\sinh(\alpha^{*}\alpha_{0}+\alpha\alpha_{0})-i\frac{1}{\sqrt{2}}\sinh(\alpha^{*}\alpha_{0}-\alpha\alpha_{0})\}
\end{array}$ & $k>2$, $k$ even\tabularnewline
\hline 
\textcolor{black}{$t=\pi/\Omega$} & \textcolor{black}{$\frac{1}{\pi}\exp(-|\alpha|^{2}-|\alpha_{0}|^{2}-\alpha_{0}(\alpha^{*}+\alpha))$} & \textcolor{black}{all $k$}\tabularnewline
\hline 
\end{tabular}
\par\end{centering}
\caption{\textcolor{black}{\label{tab:Q_alpha}$Q(\alpha)$ functions evaluated
for various times and $k$.}}

\textcolor{blue}{}
\end{table}

\twocolumngrid


\begin{thebibliography}{10}
\bibitem{s-cat}\foreignlanguage{australian}{E. Schroedinger, ``The
Present Status of Quantum Mechanics'', Die Naturwissenschaften. \textbf{23},
807 (1935). \textcolor{red}{}}

\bibitem{cat-states-review}\foreignlanguage{australian}{Florian Fröwis,
Pavel Sekatski, Wolfgang Dür, Nicolas Gisin, and Nicolas Sangouard,
``Macroscopic quantum states: measures, fragility, and implementations'',
Rev. Mod. Phys. \textbf{90}, 025004 (2018) }

\bibitem{cat-states}S. Haroche, ``Nobel Lecture: Controlling photons
in a box and exploring the quantum to classical boundary'', Rev.
Mod. Phys. \textbf{85}, 1083 (2013). D. J. Wineland, ``Nobel Lecture:
Superposition, entanglement, and raising Schrödinger\textquoteright s
cat'', Rev. Mod. Phys. \textbf{85}, 1103 (2013). 

\bibitem{supercond-microwave-cats}B. Vlastakis et al., Science \textbf{342},
607 (2013). C. Wang et al., Science \textbf{352}, 1087 (2016).

\bibitem{collapse-revival-bec}Markus Greiner, Olaf Mandel, Theodor
Hånsch and Immanuel Bloch, \textquotedblleft Collapse and revival
of the matter wave field of a Bose-Einstein condensate\textquotedblright ,
Nature \textbf{419}, 51 (2002).

\bibitem{collapse-revival-super-circuit}Gerhard Kirchmair et al.,
``Observation of the quantum state collapse and revival due to a
single-photon Kerr effect\textquotedblright , Nature \textbf{495},
205 (2013). 

\bibitem{legggarg}A. Leggett and A. Garg, \textquotedblleft Quantum
mechanics versus macroscopic realism: is the flux there when nobody
looks?\textquotedblright , Phys. Rev. Lett. \textbf{54}, 857 (1985).

\bibitem{bell-1}\foreignlanguage{australian}{J. S. Bell, ``On the
Einstein-Podolsky-Rosen Paradox'', Physics \textbf{1,} 195 (1964).}

\bibitem{emeryreview}\foreignlanguage{australian}{C. Emary, N. Lambert
and F. Nori, ``Leggett-Garg inequalities'', Rep. Prog. Phys \textbf{77},
016001 (2014).\textcolor{red}{}}

\bibitem{small-systemlg}\foreignlanguage{australian}{J. Dressel et
al., ''Experimental Violation of Two-Party Leggett-Garg Inequalities
with Semiweak Measurements'', Phys. Rev. Lett. \textbf{106}, 040402
(2011). J. S. Xu et al., ``Experimental violation of the Leggett-Garg
inequality under decoherence'', Scientific Report \textbf{1}, 101
(2011).\textcolor{blue}{{} }V. Athalye, S. S. Roy, and T. S. Mahesh,
``Investigation of the Leggett-Garg Inequality for Precessing Nuclear
Spins'', Phys. Rev. Lett. \textbf{107}, 130402 (2011). A. M. Souza,
I. S. Oliveira and R. S. Sarthour, ``A scattering quantum circuit
for measuring Bell's time inequality: a nuclear magnetic resonance
demonstration using maximally mixed states'', New J. Phys. \textbf{13}
053023 (2011). G. Waldherr et al.,``Violation of a Temporal Bell
Inequality for Single Spins in a Diamond Defect Center'', Phys.Rev.
Lett. \textbf{107,} 090401 (2011). H. Katiyar et al., ``Violation
of entropic Leggett-Garg inequality in nuclear spins'', Phys. Rev.
A \textbf{87}, 052102\textcolor{blue}{{} }\textcolor{black}{(2013).
}R. E. George et al., ``Opening up three quantum boxes causes classically
undetectable wavefunction collapse'', Proc. Natl Acad. Sci. \textbf{110}
3777 (2013). G. C. Knee et al., ``Violation of a Leggett-{}-Garg
inequality with ideal non-invasive measurements'', Nature Commun.
\textbf{3}, 606 (2012).\textcolor{red}{} \textcolor{red}{}\textcolor{blue}{}}

\selectlanguage{australian}%
\bibitem{lgexpphotonweak}M. E. Goggin, et al., ``Violation of the
Leggett-Garg inequality with weak measurements of photons'', Proc.
Natl. Acad. Sci. \textbf{\textcolor{black}{108}}, 1256 (2011).

\selectlanguage{english}%
\bibitem{stat-1zhou}\foreignlanguage{australian}{Z. Q. Zhou, S. Huelga,
C-F Li and G-C Guo, ``Experimental Detection of Quantum Coherent
Evolution through the Violation of Leggett-Garg-Type Inequalities'',
Phys. Rev. Lett. \textbf{115}, 113002 (2015). }

\bibitem{stat-2-neutrino}\foreignlanguage{australian}{J. A. Formaggio
et al., ``Violation of the Leggett-Garg Inequality in Neutrino Oscillations'',
Phys. Rev. Lett. \textbf{117}, 050402 (2016).}

\bibitem{experiment-lg}\foreignlanguage{australian}{A. Palacios-Laloy,
F. Mallet, F. Nguyen, P. Bertet, Denis Vion, Daniel Esteve and Alexander
N. Korotkov, ``Experimental violation of a Bell's inequality in time
with weak measurement'', Nature Phys. \textbf{6,} 442 (2010).}

\bibitem{atom-lg}\foreignlanguage{australian}{ C. Robens, W. Alt,
D. Meschede, C. Emary, and A. Alberti, ``Ideal Negative Measurements
in Quantum Walks Disprove Theories Based on Classical Trajectories'',
Phys. Rev. X \textbf{5}, 011003 (2015). }

\selectlanguage{australian}%
\bibitem{NSTmunro}\textcolor{black}{G. C. Knee, }K. Kakuyanagi, M.-C.
Yeh, Y. Matsuzaki, H. Toida, H. Yamaguchi, S. Saito, A. J. Leggett
and W. J. Munro, ``A strict experimental test of macroscopic realism
in a superconducting flux qubit'', Nat. Commun. \textbf{7}, 13253
(2016).

\bibitem{jordan_kickedqndlg2} A. N. Jordan, A. N. Korotkov, and M.
Buttiker, ``Leggett-Garg Inequality with a Kicked Quantum Pump'',
Phys. Rev. Lett.\textbf{ 97}, 026805 (2006). 

\bibitem{massiveosci}A. Asadian, C. Brukner and P. Rabl, ``Probing
Macroscopic Realism via Ramsey Correlation Measurements'', Phys.
Rev. Lett. \textbf{112}, 190402 (2014).

\bibitem{Mitchell}C. Budroni, G. Vitagliano, G. Colangelo, R. J.
Sewell, O. Gühne, G. Tóth, and M. W. Mitchell, ``Quantum Nondemolition
Measurement Enables Macroscopic Leggett-Garg Tests'', Phys. Rev.
Lett. \textbf{115}, 200403 (2015).\textcolor{red}{}

\selectlanguage{english}%
\bibitem{bogdan-two-well}\foreignlanguage{australian}{B. Opanchuk,
L. Rosales-Zárate, R. Y. Teh, and M. D. Reid, ``Quantifying the mesoscopic
quantum coherence of approximate NOON states and spin-squeezed two-mode
Bose-Einstein condensates'', Phys. Rev. A \textbf{94}, 062125 (2016); }

\bibitem{laura-lg-two-well}\foreignlanguage{australian}{L. Rosales-Zárate,
B. Opanchuk, Q. Y. He, and M. D. Reid, ``Leggett-Garg tests of macrorealism
for bosonic systems including two-well Bose-Einstein condensates and
atom interferometers'', Phys. Rev. A \textbf{97}, 042114 (2018). }

\bibitem{milburn-holmes}G. Milburn and C. Holmes, \textquotedblleft Dissipative
quantum and classical Liouville mechanics of the anharmonic oscillator\textquotedblright ,
Phys. Rev. Lett. \textbf{56}, 2237 (1986). 

\bibitem{yurke-stoler}B. Yurke and D. Stoler, \textquotedblleft Generating
quantum mechanical superpositions of macroscopically distinguishable
states via amplitude dispersion\textquotedblright , Phys. Rev. Lett.
\textbf{57}, 13 (1986).

\bibitem{wrigth-walls-gar}E. Wright, D. Walls and J. Garrison, \textquotedblleft Collapses
and revivals of Bose-Einstein condensates formed in small atomic samples\textquotedblright ,
Phys. Rev. Lett. \textbf{77}, 2158 (1996). 

\bibitem{triple-non}E. A. Rojas Gonz\'{ }alez, A. Borne, B. Boulanger,
J. A. Levenson, and K. Bencheikh, ``Continuous variable triple-photon
states quantum entanglement'', Phys. Rev. Lett. \textbf{\textcolor{black}{120}},
043601 (2018).\textcolor{red}{}

\selectlanguage{australian}%
\bibitem{weak} Y. Aharonov, \textcolor{red}{} D. Albert and L. Vaidmann,\textcolor{blue}{{}
}\textcolor{black}{``How the result of a measurement of a component
of the spin of a spin-1/2 particle can turn out to be 100'', Phys.
Rev. Lett. }\textbf{\textcolor{black}{60}}\textcolor{black}{, 1351
(1988). N. S. Williams, A. N. Jordan,  ``Weak Values and the Leggett-Garg
Inequality in Solid-State Qubits'', Phys. Rev. Lett. }\textbf{\textcolor{black}{100}}\textcolor{black}{,
026804 (2008).}

\bibitem{weakLGbellexp_review} J. Dressel, M. Malik, F. M. Miatto,
A. N. Jordan and R. W. Boyd, ``Understanding quantum weak values:
Basics and applications'', Rev. Mod. Phys. \textbf{86}, 307 (2014).\textcolor{red}{{}
} T.\LyXThinSpace C. White et al., ``Preserving entanglement during
weak measurement demonstrated with a violation of the Bell-Leggett-Garg
inequality'', NPJ Quantum Inf. \textbf{2}, 15022 (2016). B. L. Higgins,
M. S. Palsson, G. Y. Xiang, H. M. Wiseman, and G. J. Pryde, ``Using
weak values to experimentally determine negative probabilities in
a two-photon state with Bell correlations'', Phys. Rev. A \textbf{91},
012113 (2015).

\selectlanguage{english}%
\bibitem{weak-noon}\foreignlanguage{australian}{L. Rosales-Zárate,
B. Opanchuk, and M. D. Reid, ``Weak measurements and quantum weak
values for NOON states'', Phys. Rev. A \textbf{97}, 032123 (2018).}

\bibitem{bell-review}\foreignlanguage{australian}{Clauser and Shimony,
``Bell's theorem: experimental tests and implications'', Rep. Prog.
Phys. \textbf{41}, 1881 (1978).}

\bibitem{Husimi-Q}Kôdi Husimi (1940). \textquotedbl{}Some Formal
Properties of the Density Matrix\textquotedbl{}, Proc. Phys. Math.
Soc. Jpn. 22: 264-314. 
\end{thebibliography}
\end{document}